\documentclass[useAMS,usenatbib]{mn2e}
\usepackage{amsmath}
\usepackage{multicol}
\usepackage{epsfig}
\usepackage{rotating}
\usepackage[usenames, dvipsnames]{color}

\title[M. E. Bell et al]
{The Murchison Widefield Array Transients Survey (MWATS).  A search for low frequency variability in a bright Southern hemisphere sample}
\author[M.E.Bell]{M. E. Bell$^{1, 2, 3}$\thanks{E-mail:
martin.bell@uts.edu.au (MEB)}, 
Tara Murphy$^{3,4}$, P. J. Hancock$^{3,8}$, J.~R.~Callingham$^{10}$, S. Johnston$^{2}$, 
\newauthor D.~L. Kaplan$^{5}$, R. W. Hunstead$^{4}$, E.~M. Sadler$^{3, 4}$,  S. Croft$^{6,7}$, S.~V.  White$^{8}$, \and 
N. Hurley-Walker$^{8}$,
R. Chhetri$^{3,8}$, 
J. S. Morgan$^{8}$, 
P. G. Edwards$^{2}$, 
A. Rowlinson$^{9,10}$,
\and 
A. R. Offringa$^{10}$, 
G.~Bernardi$^{11,12,13}$, 
J.~D.~Bowman$^{14}$, 
F.~Briggs$^{15}$, 
R.~J.~Cappallo$^{16}$, \and
A.~A.~Deshpande$^{17}$, 
B.~M.~Gaensler$^{3,4,18}$, 
L.~J.~Greenhill$^{13}$, 
B.~J.~Hazelton$^{19}$, \and 
M.~Johnston-Hollitt$^{20}$, 
C.~J.~Lonsdale$^{16}$,
S.~R.~McWhirter$^{16}$, 
D.~A.~Mitchell$^{2,3}$, \and 
M.~F.~Morales$^{19}$,  
E.~Morgan$^{6}$, 
D.~Oberoi$^{20}$, 
S.~M.~Ord$^{2,3}$, 
T.~Prabu$^{17}$, 
N.~Udaya~Shankar$^{17}$, \and
K.~S.~Srivani$^{17}$, 
R.~Subrahmanyan$^{3,17}$, 
S. J. Tingay$^{3,8}$, 
R.~B.~Wayth$^{3,8}$, 
R.~L.~Webster$^{3,22}$, \and
A.~Williams$^{8}$, 
C.~L.~Williams$^{16}$ \\
\\
$^{1}$ University of Technology Sydney, 15 Broadway, Ultimo NSW 2007, Australia\\
$^{2}$ CSIRO Astronomy and Space Science, PO Box 76, Epping NSW 1710, Australia\\
$^{3}$ ARC Centre of Excellence for All-sky Astrophysics (CAASTRO)\\
$^{4}$ Sydney Institute for Astronomy (SIfA), School of Physics, The University of Sydney, NSW 2006, Australia \\
$^{5}$ Department of Physics, University of Wisconsin-Milwaukee, 1900 E. Kenwood Boulevard, Milwaukee, WI 53211, USA \\
$^{6}$ University of California, Berkeley, Astronomy Dept., 501 Campbell Hall \#3411, Berkeley, CA 94720, USA \\
$^{7}$ Eureka Scientific, Inc., 2452 Delmer Street Suite 100, Oakland, CA 94602, USA \\
$^{8}$ International Centre for Radio Astronomy Research, Curtin University, Bentley, WA 6845, Australia\\
$^{9}$ Anton Pannekoek Institute, University of Amsterdam, Postbus 94249, 1090 GE, Amsterdam, The Netherlands\\
$^{10}$ ASTRON, Netherlands Institute for Radio Astronomy, Postbus 2, 7990 AA Dwingeloo, The Netherlands \\
$^{11}$ SKA SA, 3rd Floor, The Park, Park Road, Pinelands, 7405, South Africa \\
$^{12}$ Department of Physics and Electronics, Rhodes University, PO Box 94, Grahamstown 6140, South Africa\\
$^{13}$ Harvard-Smithsonian Center for Astrophysics, Cambridge, MA 02138, USA\\
$^{14}$ School of Earth and Space Exploration, Arizona State University, Tempe, AZ 85287, USA\\
$^{15}$ Research School of Astronomy and Astrophysics, Australian National University, Canberra, ACT 2611, Australia\\
$^{16}$ MIT Haystack Observatory, Westford, MA 01886, USA\\
$^{17}$ Raman Research Institute, Bangalore 560080, India\\
$^{18}$ Dunlap Institute for Astronomy and Astrophysics, University of Toronto, ON, M5S 3H4, Canada\\
$^{19}$ Department of Physics, University of Washington, Seattle, WA 98195, USA\\
$^{20}$ School of Chemical \& Physical Sciences, Victoria University of Wellington, PO Box 600, Wellington 6140, New Zealand\\
$^{21}$ National Centre for Radio Astrophysics, Tata Institute for Fundamental Research, Pune 411007, India\\
$^{22}$ School of Physics, The University of Melbourne, Parkville, VIC 3010, Australia}

\begin{document}
\pagerange{\pageref{firstpage}--\pageref{lastpage}} \pubyear{2002}

\maketitle
\label{firstpage}
\begin{abstract}
We report on a search for low-frequency radio variability in 944 bright ($>4$\,Jy at 154\,MHz) unresolved, extragalactic radio sources monitored monthly for several years with the Murchison Widefield Array. In the majority of sources we find very low levels of variability with typical modulation indices $<5$\%. We detect 15 candidate low frequency variables that show significant long term variability ($>$2.8 years) with time-averaged modulation indices $\overline M=3.1-7.1$\%. With 7/15 of these variable sources having peaked spectral energy distributions, and only 5.7\% of the overall sample having peaked spectra, we find an increase in the prevalence of variability in this spectral class. We conclude that the variability seen in this survey is most probably a consequence of refractive interstellar scintillation and that these objects must have the majority of their flux density contained within angular diameters less than 50 milli-arcsec (which we support with multi-wavelength data). At 154~MHz we demonstrate that interstellar scintillation time-scales become long ($\sim$decades) and have low modulation indices, whilst synchrotron driven variability can only produce dynamic changes on time-scales of hundreds of years, with flux density changes less than one milli-jansky (without relativistic boosting). From this work we infer that the low frequency extragalactic southern sky, as seen by SKA-Low, will be non-variable on time-scales shorter than one year.   
\end{abstract}


\section{Introduction}
The phenomenon of low frequency variability (LFV) was first clearly confirmed by \cite{Hunstead_72}. In that study, four bright extragalactic variable radio sources were identified that showed amplitude modulation of up to 20\% on time-scales under five years at 408\,MHz. At that time the variations were difficult to reconcile as they broke the Compton catastrophe limit \citep{Kell1969}, meaning that the mechanism for variability could not be intrinsic to the source. Further studies at 408\,MHz confirmed that the phenomenon was common, with 25\% of compact sources and 51\% of flat spectrum sources displaying dynamic properties (\citealt{McAdam_79}; \citealt{Fanti_81}; \citealt{spangler_1989}; see also \citealt{Jauncey_2016} for a review).      

New low-frequency interferometers such as the Murchison Widefield Array \citep[MWA;][]{MWA} and the Low Frequency Array \citep[LOFAR;][]{LOFAR} allow for greatly improved spatial, temporal and spectral coverage. Hence we can now efficiently probe large samples of low-frequency sources for long-timescale dynamic changes. 

Scaling up the \cite{Hunstead_72} results to the abundance of radio sources now routinely being detected by these new instruments (hundreds of thousands), one might assume the extragalactic low frequency sky to be highly variable. In this paper we conduct a comprehensive analysis of the abundance of low-frequency variables as seen by the MWA and on the basis of this make predictions for the Square Kilometre Array \citep{SKA-Low}. 

The mechanism driving the variability in the sources identified by \citet{Hunstead_72} has been determined to be interstellar scintillation (ISS) caused by the turbulent and ionised interstellar medium (ISM; see \citealt{Rickett_84}). Scintillation can be divided into several different subclasses that are a function of source angular size, observing frequency and line of sight through the ISM which will influence the amplitude  and time-scale of variability (see \citealt{Lazio_2004} for a review). Specifically, refractive scintillation is bending of different ray paths through the clumpy ISM, while diffractive scintillation is interference between different ray paths. 

LFV is now attributed to refractive interstellar scintillation (RISS) and is prominent when the source sizes are $\sim$1\,milli-arcsecond across and the observing frequency is low $<500$\,MHz \citep{Rickett_86}. At higher frequencies intra-day variability (IDV; \citealt{J1819}; \citealt{Bignall_2003}) can be thought of as the higher-frequency ``sibling" propagation effect to LFV. IDV can cause large intensity fluctuations on short time-scales ($<24$\,hours) and is most prominent at frequencies $\sim5$\,GHz. Sweeping from high (GHz) to low (MHz) frequencies we transition from the weak scattering regime (less than one radian of phase variation, with modulation index $<1$)  to the strong ($\gg 1$\,radian of phase variation).  In the strong regime if the source is compact enough compared to an isoplanatic patch (typically only for pulsars)  it will exhibit diffractive scintillation, but larger sources such as active galaxies exhibit refractive scintillation with slow (time-scale $\propto \nu^{-2.2}$) lower-amplitude (modulation index $\propto \nu^{0.57}$) fluctuations  \citep{Lazio_2004}. 

Interplanetary Scintillation (IPS; \citealt{Hewish_1964}) requires a contrast in density of the intervening medium, this time in the interplanetary medium, which is driven by inhomogeneities in the solar wind. Although IPS is typically observed at small solar elongations it has been detected serendipitously by the MWA in night-time observing \citep{kaplan_2015} with flux density variations measured on the time-scale of seconds.  This has been further exploited by \cite{Morgan_2017} and \cite{Rajan} to catalogue hundreds of sources with sub-arcsec structure across the sky. A major finding of this work is that almost all peaked-spectrum sources show strong IPS, meaning that they are compact on sub-arcsecond scales.

A number of surveys have now studied the time-domain radio sky specifically searching for (and studying) propagation effects at a range of frequencies. For example, at GHz frequencies, the Micro-Arcsecond Scintillation-Induced Variability survey (MASIV; \citealt{Lovell_2003}; \citealt{Lovell_2008}; \citealt{MASIVIII}) found that over 50\% of flat spectrum sources show significant variability at 5\,GHz. The authors do report, however, that a higher abundance of these variables are detected in their weak sample ($\sim$0.1~Jy) as opposed to their strong sample ($\sim$1~Jy).   
The faster, more intense variability seen at 5\,GHz (as discussed above) is driven by weak ISS (see \citealt{Jauncey_2016} for a review). 

At frequencies comparable to this work (154\,MHz), some observational studies have reported that the prevalence of LFV is more rare than at 408\,MHz 
For example, \cite{McGilchrist} show that only 1.1\% of a sample of 811 sources with flux densities $>0.3$~Jy show fractional variability $>$15\% on time-scales of one year at 151~MHz (also see \citealt{Riley_1993}). In contrast, \cite{Slee_1988} report that 13.3\% of a sample of 412  sources $>1$~Jy show fractional variability $>15$\% at 160~MHz. The total length of the \cite{Slee_1988} study was, however, longer (up to 14 years), although variability was still reported on month time-scales. In both cases, however, the abundance of variables is lower  than reported at 408\,MHz ($\sim25$\%; see \citealt{McAdam_79} and \citealt{Fanti_81}). 

In more recent work, \cite{Rowlinson_2016} probed an unbiased sample of predominantly extragalactic radio sources with flux densities $\geq 0.5$~Jy over a period of one year with the MWA at 182~MHz. In this study, over 100 hours of observations were analysed to produce more than ten-thousand wide-field images (600\,deg$^2$ each) containing around 5000 sources. For this sample, low-levels of variability were reported with modulation indices typically below 10\% \citep{Rowlinson_2016}. In comparison, \cite{Bell_2014} used the pathfinder MWA instrument to study LFV and found 1.5\% of sources with flux densities $\geq 3$~Jy showed significantly fractional variability with variations $>10$\% on time-scales of one year. 

Clearly the abundance of variable sources is lower at low frequencies, which is supported by the theoretical frequency dependence of RISS. But with the recent renaissance in low-frequency radio astronomy the time was right to revisit the field of LFV and pursue a systematic census of its abundance prior to SKA-Low operations. The MWA also has exceptional spectral coverage at low frequencies, so for the first time we could explore the detailed spectral energy distributions of variables. We focussed this survey on a pre-selected sample of 944 compact extragalactic sources distributed over the Southern hemisphere and probed time-scales up to 2.8 years. 

This paper uses data taken for the Murchison Widefield Array Transients Survey (MWATS: P.I.\ M.~Bell). MWATS was an observing campaign aimed at sampling the dynamic radio sky on time-scales of minutes to years over the Southern hemisphere at 154\,MHz. MWATS has already provided time-domain measurements of objects such as pulsars \citep{Bell2016}; exoplanets \citep{Murphy_2015}; the ionosphere (\citealt{Cleo_PB}; \citealt{Cleo_2015b}; \citealt{Cleo_2016}) and low-mass stars. The end goal of MWATS was to provide a complete time-domain Southern hemisphere census of known and previously unknown objects (i.e.\ transients) prior to SKA-Low operations. 

In this paper we revisit the topic of low frequency variability through the lens of our survey(s). In $\S$\ref{sec:survey} we describe the survey specifications and sample selection. In $\S$3 we describe the calibration, imaging and light-curve extraction procedure. In $\S$4 we examine the time-domain, spectral and multi-wavelength properties of these objects to identify possible variables. In $\S$5 we discuss the physical interpretation of our results and compare them with past and upcoming surveys. Unless otherwise quoted, all uncertainties correspond to $1\sigma$ confidence. We define the relationship between flux density $S_{\nu}$, frequency $\nu$ and spectral index $\alpha$ as $S_{\nu} \propto \nu^\alpha$. 

\section{Survey specifications}
\label{sec:survey}
The observations for this survey commenced on 2013 September 16 and ended on 2016 June 25, covering a time span of 2.8 years. In total there were 24 nights of data, each separated by more than 10 days, giving a total of 5639 images. Each image covered an area of sky approximately 1000\,deg$^{2}$. Observations were centred at 154\,MHz with an observing bandwidth of 30.72\,MHz. The bandwidth was divided into 768 channels of width 40\,kHz. The correlator integration cycle was either 0.5 or 2\,seconds; it was increased in later observations (after 2014) to reduce disk usage.   

The survey strategy utilised drift scanning to repeatedly observe three meridian points at declination centres of $-26^{\circ}$ (zenith) $+1.6^{\circ}$ and $-55^{\circ}$ (see Figure~\ref{map}). Owing to the large field of view of the MWA this strategy gave uniform coverage at 154\,MHz between declinations of  $-70^{\circ}$ and $+20^{\circ}$. Each observation consisted of a 112 second integration at each of the meridian points in turn. The mean time difference between all observations regardless of declination was two minutes. For the same declination strip it was six minutes, due to declination switching. Observations were always obtained at night and seasonal sky rotation allowed us to cover the complete RA range per year. This had an impact on the cadence of observations for the objects in our sample. A given source was typically observable for 3$-$4 months per year depending on its proximity to the drift scan centres. A summary of the specifications is given in Table~\ref{observations}. 

\subsection{Sample selection}
Using the GaLactic and Extragalactic All-sky Murchison Widefield Array (GLEAM) survey \citep{Randall} we extracted a biased sample of bright ($>4$\,Jy at 151~MHz) and possibly compact radio sources defined via visual and spectral inspection from the Extra-Galactic Catalogue (EGC; \citealt{GLEAM}, White et al., in prep.). We take all components in the GLEAM EGC that are brighter than 4\,Jy (total of 1879 objects). From visual inspection we removed objects that had the following properties: 
\begin{enumerate}
\item were resolved into two or more components in images from the NRAO VLA Sky Survey (NVSS; \citealt{NVSS}) or the Sydney University Molonglo Sky Survey (SUMSS; \citealt{SUMMS}); 
\item were confused in the MWA beam, i.e., there were two unresolved NVSS or SUMSS sources within the error ellipse of the MWA beam.  
\end{enumerate}

\begin{figure}
\centering
\includegraphics[scale=0.73]{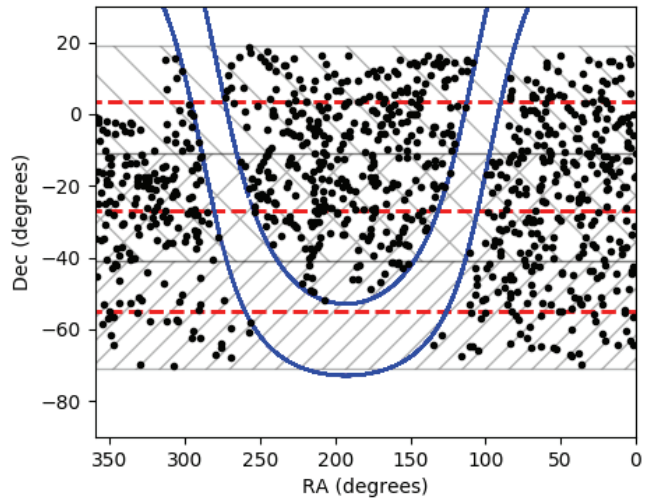}
\caption{The spatial distribution of objects in our sample. The hatched areas shown the drift scan regions which are centred on declinations of $+$1.6$^{\circ}$, $-$26$^{\circ}$ and $-$55$^{\circ}$ (red dashed lines). Sources in the Galactic plane are missing as they are not included in the GLEAM EGC \citep{GLEAM}. This exclusion zone is defined by the blue lines where $b = \pm 10^{\circ}$.}
\label{map}
\end{figure}

\noindent This left a sample of 1143 objects. \\
\indent Within our survey area we then constructed light-curves for these objects which gave a final sample of 944 objects. 
This lower sample size is due to the different footprints of MWATS and the GLEAM survey.
GLEAM covers a declination range $-90^{\circ}<\delta<+30^{\circ}$ whereas MWATS covers $-70^{\circ}<\delta<+20^{\circ}$, leading to a smaller overlap region and hence smaller sample size.  
The spatial distribution of our 944 objects is shown in Figure~\ref{map}. Note, the GLEAM EGC does not report sources that fell within a number of exclusion zones (see \citealt{GLEAM} for full details of exclusion regions). For example, sources within 10$^\circ$ of the Galactic plane, or within 5.5$^{\circ}$ and 2.5$^{\circ}$ of the Large or Small Magellanic Clouds respectively were excluded. As our sample was derived from the GLEAM EGC we consequently do not consider sources in these regions. 

\begin{table}
\centering
\caption{Properties of observations.}
\begin{tabular}{|c|c|}
\hline
Property & Value \\
\hline
Total images & 5639 \\
Centre frequency & 154\,MHz\\
Bandwidth & 30.72\,MHz\\
Channelisation & 40\,kHz\\
Integration time per image & 112\,s \\
Correlation time & 0.5\,s or 2\,s\\
Image size (pixels) & 4000 $\times$ 4000 \\
Pixel size & $0.75^{\prime} \times 0.75^{\prime} $ \\
Beam diameter (FWHM) & 2.25$^{\prime}$ \\
Briggs weighting & $-$1 \\
UV selection (k$\lambda$) & $>0.03$ k$\lambda$ \\
Declinations & $+$1.6$^{\circ}$, $-$26$^{\circ}$, $-$55$^{\circ}$ \\
Cadence & minutes, months and years  \\
Typical image noise\\ 
(extragalactic pointing) & 20~mJy/beam\\
Typical image noise\\ 
(Galactic pointing) & 100~mJy/beam\\
Survey duration & 2.8 years \\
\hline
\label{observations}
\end{tabular}
\end{table}

\section{Data reduction}
\subsection{Calibration}
For each night of observations a single 112-second observation centred on a bright, well-modelled (not necessarily unresolved) source was obtained for phase calibration purposes. The models used for calibration were extracted from previous radio surveys of the southern hemisphere, namely SUMSS or the VLA Low-frequency Sky Survey  (VLSS; \citealt{VLSS}). They are drawn from the same set of models used for calibration of the GLEAM survey. The model images were inverse Fourier transformed to generate a model set of visibilities. A single time-independent, frequency-dependent amplitude and phase calibration solution was derived from this model and compared with the calibrator observation visibilities. These gain solutions were then applied to the nightly drift scan observations prior to imaging. 

\subsection{Imaging} 
The imaging procedure was almost identical to that done in \cite{Bell2016} and very close to the processing performed for the GLEAM survey. The major change for this work (compared with \citealt{Bell2016}) was the addition of the peeling and deep {\sc clean} steps described below. For each of the observations we did the following steps: 

\begin{itemize}
\item \textit{Flagging:} The visibilities were flagged for radio frequency interference using the {\sc aoflagger} algorithm (\citealt{offringa_2012}) and converted into {\sc casa} measurement set format using the MWA preprocessing pipeline {\sc cotter}. Approximately 1\% of the visibilities were removed at this stage (see \citealt{Andre_2015} for more details). 

\item \textit{Bandpass calibration:} Phase solutions derived from the calibrator observation were applied to the visibilities (as discussed above).

\item \textit{Peeling:} Bright sources ($>$50\,Jy) that fell within the sidelobe regions were peeled and removed to aid in both self-calibration (described below) and image fidelity. Peeling involves using a model of the bright source and a self-calibration cycle to adequately subtract that source from the visibilities (see \citealt{GLEAM} for further details).

\item \textit{Shallow {\sc clean}:} The visibilities were deconvolved and {\sc clean}ed with 2000 iterations using the {\sc wsclean} algorithm \citep{offringa_2014}. The full 30.72~MHz bandwidth was imaged using the Multi-Frequency Synthesis (MFS) {\sc wsclean} algorithm. An RMS noise measurement $\sigma_{N}$ was taken from the images to ascertain an appropriate {\sc clean} threshold for post self-calibration imaging and deconvolution. 

\item \textit{Self calibration:} The {\sc clean} component model was inverse Fourier transformed for self-calibration purposes. A new set of phase and amplitude calibration solutions were derived from this model and applied to the data.

\item \textit{Final flagging:} A further round of flagging was performed to tackle any residual interference or erroneous visibilities. The same flagging procedure was applied as discussed above. Approximately 0.1\% of the remaining visibilities were flagged at this stage. 

\item \textit{Deep {\sc clean}:} The visibilities were then deconvolved and {\sc clean}ed. An image size of 4000 $\times$ 4000 with pixel width 0.75$^{\prime}$ and robust parameter of $-1$ (close to uniform weighting) was used (see Table \ref{observations} for further details). For images centred away from Galactic plane ($|b|>15^{\circ}$) we {\sc clean}ed with an unrestricted number of {\sc clean} components down to a cut off of $3 \times \sigma_{N}$. For images centred closer to the Galactic plane ($|b|<15^{\circ}$) we {\sc clean}ed more deeply with a cut off of $1 \times \sigma_{N}$. This was because there is a larger amount of diffuse material at low Galactic latitudes and more {\sc clean} components were typically required to reduce the noise levels. 

\item \textit{Primary beam correction:} A primary beam correction was applied to create  Stokes I images; see \cite{offringa_2014} and \cite{Sutinjo_2015} for further details. We also produced images in U, Q, V, XX and YY, although only the Stokes~I images were used for the analysis in this paper.  
\end{itemize}   

\subsection{Light-curve extraction}
\label{LC_extract}
All images for this survey were processed through the Variable and Slow Transients pipeline (VAST; \citealt{vast_paper}). The VAST pipeline performs the following processing on each of the images:

\begin{itemize}
\item \textit{Background estimation:} we use the {\sc bane} \citep{Hancock_2018} algorithm to provide accurate background estimation across each image. 

\item \textit{Source extraction:} we use the {\sc aegean} (version 1.9.5; \citealt{aegean}) source finding algorithm to identify all sources above a local significance of 5$\sigma$. We restricted source finding to within 15$^{\circ}$ of the meridian pointing centres. This was to minimise possible variability in flux density arising from uncertainty in the primary beam, which is more pronounced further away from the pointing centres (see \citealt{Cleo_PB} for discussion).  

\item \textit{Flux density scale correction:} we use the GLEAM EGC to correct the flux density scale of all of our images. For a given image, all extracted sources that are within 20$^\circ$ of the pointing centre and brighter than 40\,mJy are crossmatched with the GLEAM EGC. For that image we then calculated a linear fit to the GLEAM versus MWATS flux density points in log space (with zero intercept). The gradient of this log-fit provides a single multiplicative factor, which is applied to each of the images to bring them onto the required flux density scale. 

\item \textit{Positional correction:} we apply a position-dependent correction for each of the flux density measurements. As the ionosphere can produce positional offsets as a function of time (see \citealt{Cleo_PB}) we use the GLEAM EGC as a reference frame for corrections. The position corrections for an image were determined by matching the sources extracted from the image with the GLEAM EGC. The matching radius was 0.1$^\circ$. The positional offsets were modelled using a radial basis function and the model offsets were applied to all sources in the image. This is the same process as the image warping described in \citet{GLEAM}, but it is applied to the extracted sources directly.

\item \textit{Light-curve generation:} For a source at position RA and Dec, we collate all flux density measurements from all images that are within a diameter of $0.6^{\prime}$ (in RA and Dec) to form light-curves. The flux density measurements were obtained from images averaged over the full 30.72~MHz bandwidth.  
\end{itemize} 

\subsection{Variability analysis}
\subsubsection{Raw light-curves} 
Using all source flux density measurements regardless of time-scale, we calculated the modulation index $M$, which is defined as 

\begin{equation}
M = \sigma/\overline{S},
\end{equation}

\noindent where $\sigma$ is the standard deviation and $\overline{S}$ is the arithmetic mean of the flux density measurements. We refer to these data as the raw light-curves. 

\subsubsection{Time-averaged light-curves}
As we are principally concerned with long-time-scale variations, for each source we combined all measurements on time-scales less than 10 days and re-calculated a time-averaged modulation index $\overline{M}$. The value of 10~days was chosen for our averaging time-scale as it was just below the minimum time separation between observing nights. Hence this gave per-observing-night flux density measurements over our observing campaign. 

We also calculated a linear least-squares fit to the time-averaged data. From the linear fit we used the gradient $\nabla_{F}$ and fitting error $\sigma_{F}$ to calculate a significance of the gradient $\nabla_{S} = \nabla_{F} / \sigma_{F}$. We identified candidate variable objects as having $|\nabla_{S}|>3$ and $\overline{M}>3$\%. We assessed the appropriateness of a simple linear fit by calculating the $\chi^{2}$ goodness-of-fit parameter using (a) a flat model (mean of the flux density measurements); (b) the linear fit (discussed above); and (c) a 2nd order polynomial fit. Using these values we calculated a p-value using an F-test \citep{F-test} comparing the linear fit to the flat model, which we refer to as P-slope. We also compared the polynomial fit to the linear fit, which we refer to as P-quad. A large value of P-slope (e.g. 0.99) would show that the linear fit to the data is the preferred model when compared with the flat model. A large value of P-quad would show a preference for a 2nd order polynomial fit. 

\section{Results}
\subsection{Raw light-curves}
\begin{figure}
\centering
\includegraphics[scale=0.72]{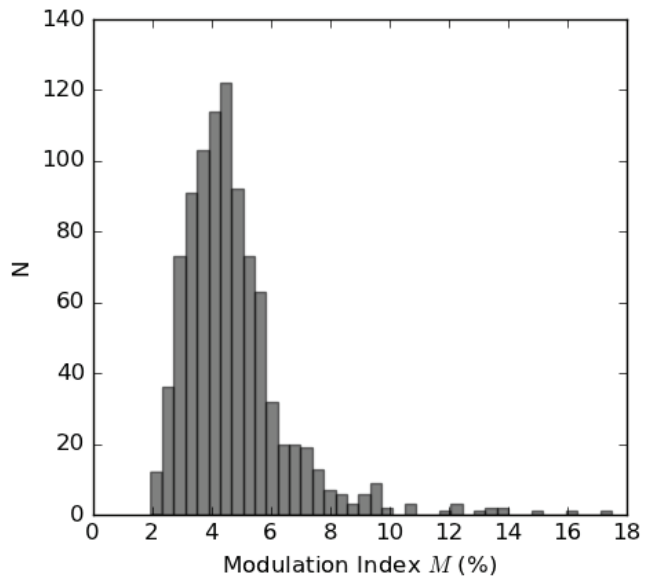}
\caption{A histogram of the modulation indices of all raw light curves. We find a median modulation index of $M=4.4 \pm 1.8$\%.}
\label{non_avg}
\end{figure}

Figure \ref{non_avg} shows the modulation index $M$ for all raw light-curves. 
As these light curves had not been averaged in time they were more susceptible to short-timescale instrumental, measurement or calibration error. Visual inspection of the small number of high-$M$ light curves ($M >$ 10\%) were characterised by the presence of erroneous measurements. Investigation of these points revealed that they were usually associated with a poor quality image or an incorrect flux density correction, resulting in an artificially large modulation index. We do not therefore believe the tail of this distribution to be caused by physically real variability. Despite these outlier cases, we still found remarkably low levels of variability in the majority of sources (on short time-scales) with a median modulation index of $M=4.4 \pm 1.8$\%. 

\subsection{Time-averaged light-curves}
\label{LC_section}
\begin{figure*}
\centering
\includegraphics[scale=0.90]{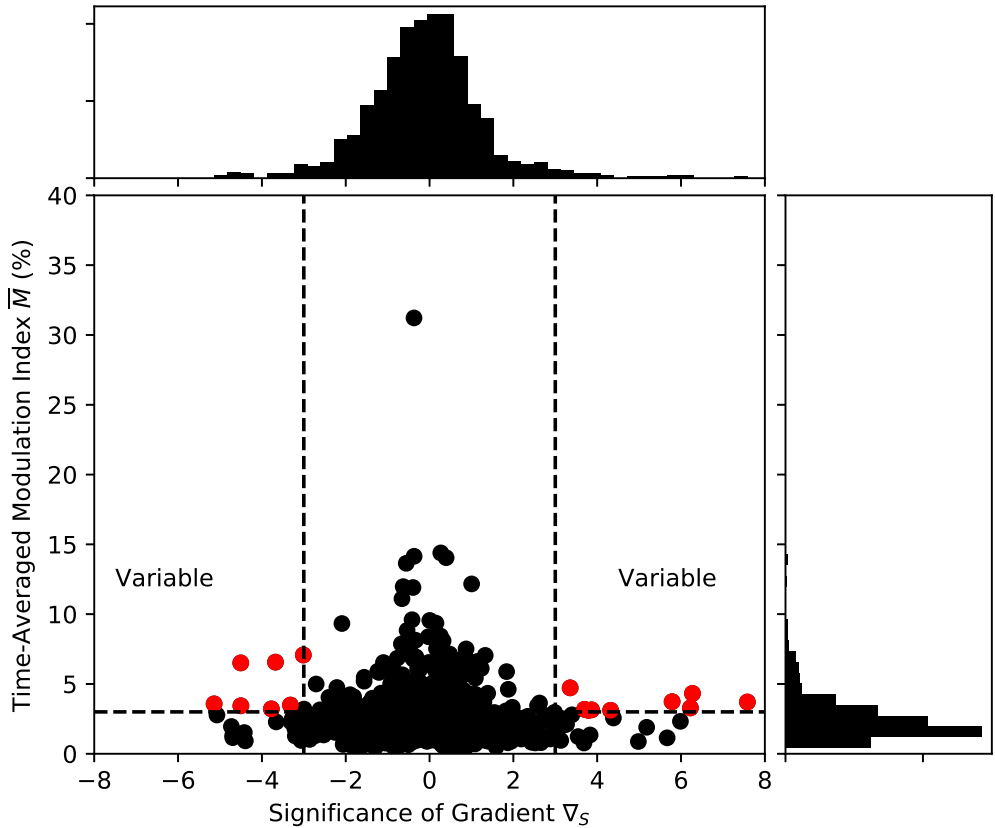}
\caption{A plot of averaged modulation index $\overline{M}$ versus significance of gradient ($\nabla_{S}$) for all sources in our sample. We find very low levels of long-term variability with a median $\overline{M} = 2.1\pm2.0$\%. Most sources remain statistically flat over our measurement time-scale (i.e. $|\nabla_{S}|<3$). Red circles denote the candidate variable sources that have a significant gradient $|\nabla_{S}|>3$ and a time-averaged modulation index $\overline{M}>3\%$.}
\label{averaged_hist}
\end{figure*}

Figure~\ref{averaged_hist} shows a histogram (right panel) of the modulation indices formed from the time-averaged light-curves. We find a median $\overline{M}$ of $2.1\pm2.0$\%, which again is very low. The top panel shows a histogram of the significance of the gradients $\nabla_{S}$ of the light-curves. In general the light-curves are exceptionally flat with a mean gradient of $\langle {\nabla}_{F} \rangle = -0.22 \pm 3.2 \times 10^{-4}$ Jy~per day.  

Examining the high-$\overline{M}$ light-curves again revealed a small number of erroneous measurements associated with a particular epoch (or night), which had driven up the average modulation index. We could have algorithmically filtered these points out, but with the small number of cases it was easy to visually inspect them and rule them out as being instrumental. 

The interesting part of the parameter space is the region where sources have a significant gradient $|\nabla_{S}|>3$ and also an average modulation index $\overline{M}>3\%$ (see Figure \ref{averaged_hist}). Within this region we find a set of 15 plausible candidate variables. These are discussed in $\S$\ref{var_notes} with the light curves and spectral energy distributions (SEDs) shown in Figures \ref{Lightcurves}--\ref{Lightcurves_5}. On the upper panel of the light curves we show the raw flux density measurements plotted sequentially. The lower panel shows the flux density as a function of calender date, with the linear fit (red line). These results are also summarised in Table~\ref{var_table}. 

For the majority of these sources (13/15) P-slope $>$ 0.90, meaning that a linear fit is preferred. Two of the sources (GLEAM J041022$-$523247 and GLEAM J102820+151129) have P-quad $>$ 0.95 (and both greater than P-slope) meaning that a quadratic fit is more appropriate. 

The SEDs (right of the light-curves) show the 20 point GLEAM EGC spectrum with red circles together with their corresponding errors. An orange line denotes a single power-law fit to all data points. If the objects are identified as having spectral curvature in Callingham et al. (2017) then a black line represents the best fit to the data using the fitting method described in Callingham et al. (2015). For GLEAM J224704$-$365746 and GLEAM J010838$+$013454, there is obvious spectral flattening above the GLEAM band that corresponds to the emission from the source becoming core-dominated. We fit these sources with a broken power-law, with the higher frequency power-law displayed as a dashed-line. Archival measurements are represented as follows:

\begin{enumerate}
\item Purple upward-pointing triangle: 74 MHz, VLA Low-frequency Sky Survey redux (VLSSr; \citealt{VLSSr});
\item Blue square: 150 MHz TIFR GMRT Sky Survey (TGSS-ADR1; \citealt{TGSS})
\item Green leftward-pointing triangle: 408~MHz, Molonglo Reference Catalogue (MRC; \citealt{MRC}); 
\item Brown rightward-pointing triangle: 843~MHz, SUMSS;
\item Navy downward-pointing triangle: 1.4 GHz, NVSS;
\item Green diamond: 4.8, 8.8 and 20~GHz, Australia Telescope 20 GHz survey (AT20G; \citealt{AT20G}).
\end{enumerate}

\begin{figure}
\centering
\includegraphics[scale=0.82]{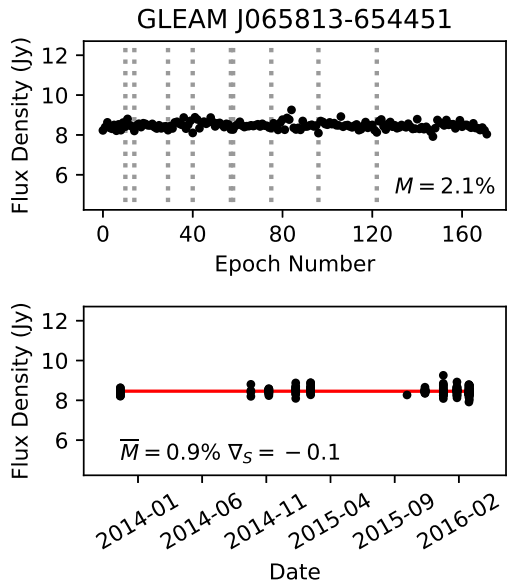}
\caption{Example of a non-variable source.}
\label{non_var}
\end{figure} 

We also made use of Very Long Baseline Interferometry (VLBI) measurements, if available. This was to ascertain the compact nature of the objects. In particular, we made use of observations from the VLBI Space Observatory Programme (VSOP; \citealt{VSOP}), the International Celestial Reference System (ICRF; \citealt{ICRF}) and the Very Long Baseline Array (VLBA; \citealt{VLBA}).

\begin{figure*}
\centering
\includegraphics[scale=0.7]{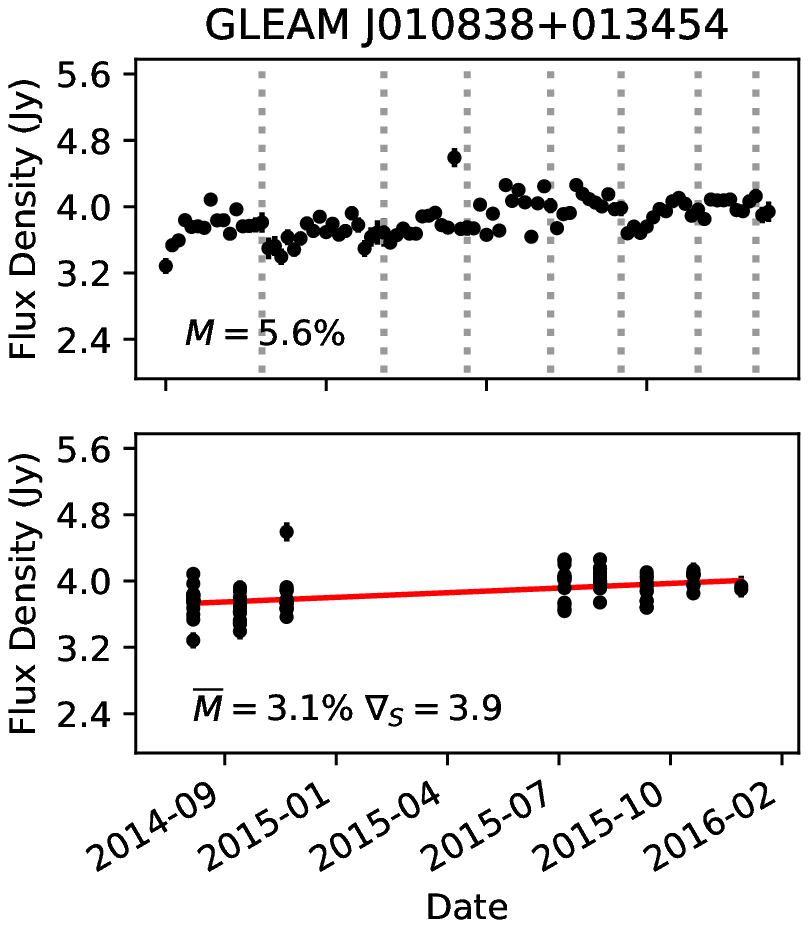}
\includegraphics[scale=0.32]{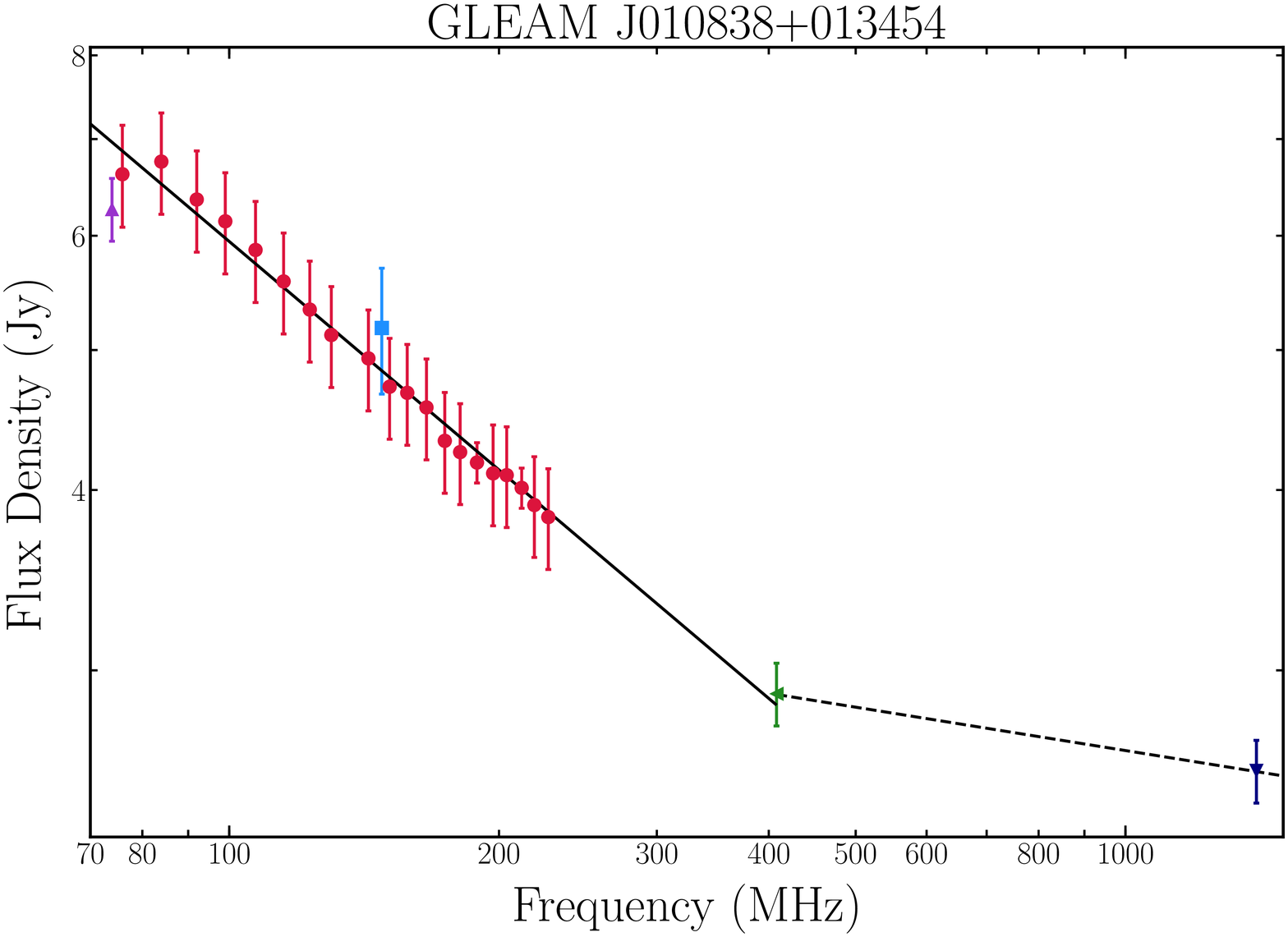}
\includegraphics[scale=0.7]{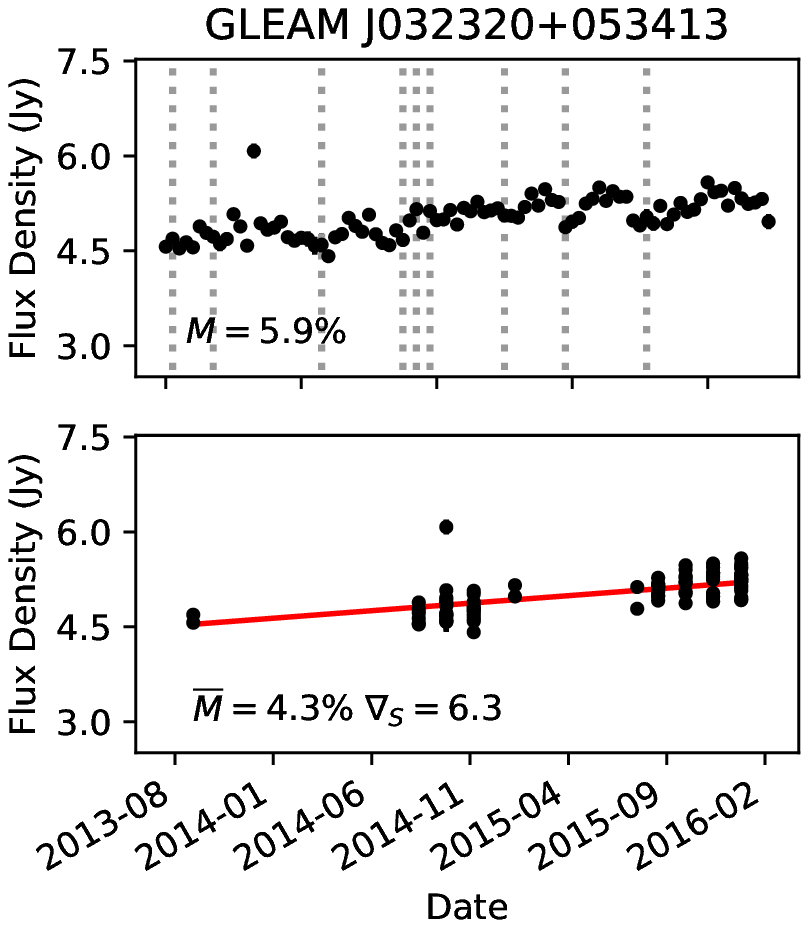}
\includegraphics[scale=0.32]{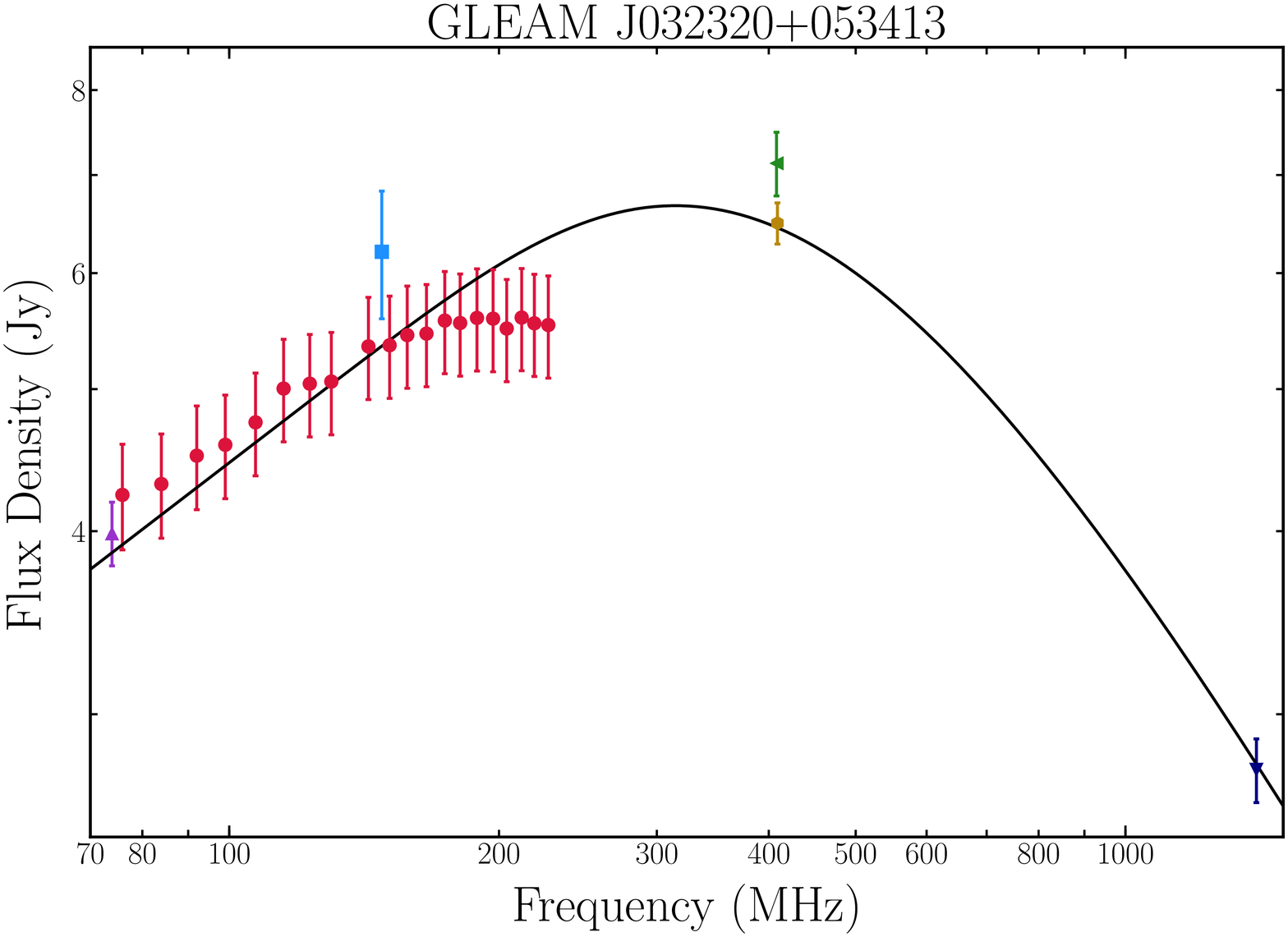}
\includegraphics[scale=0.7]{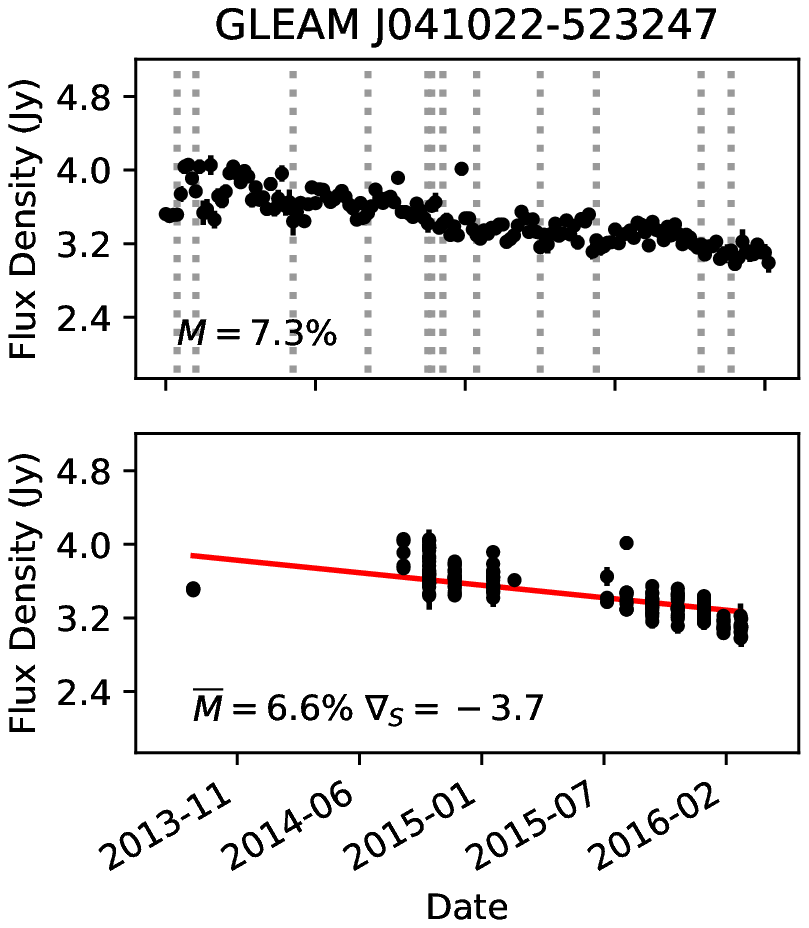}
\includegraphics[scale=0.32]{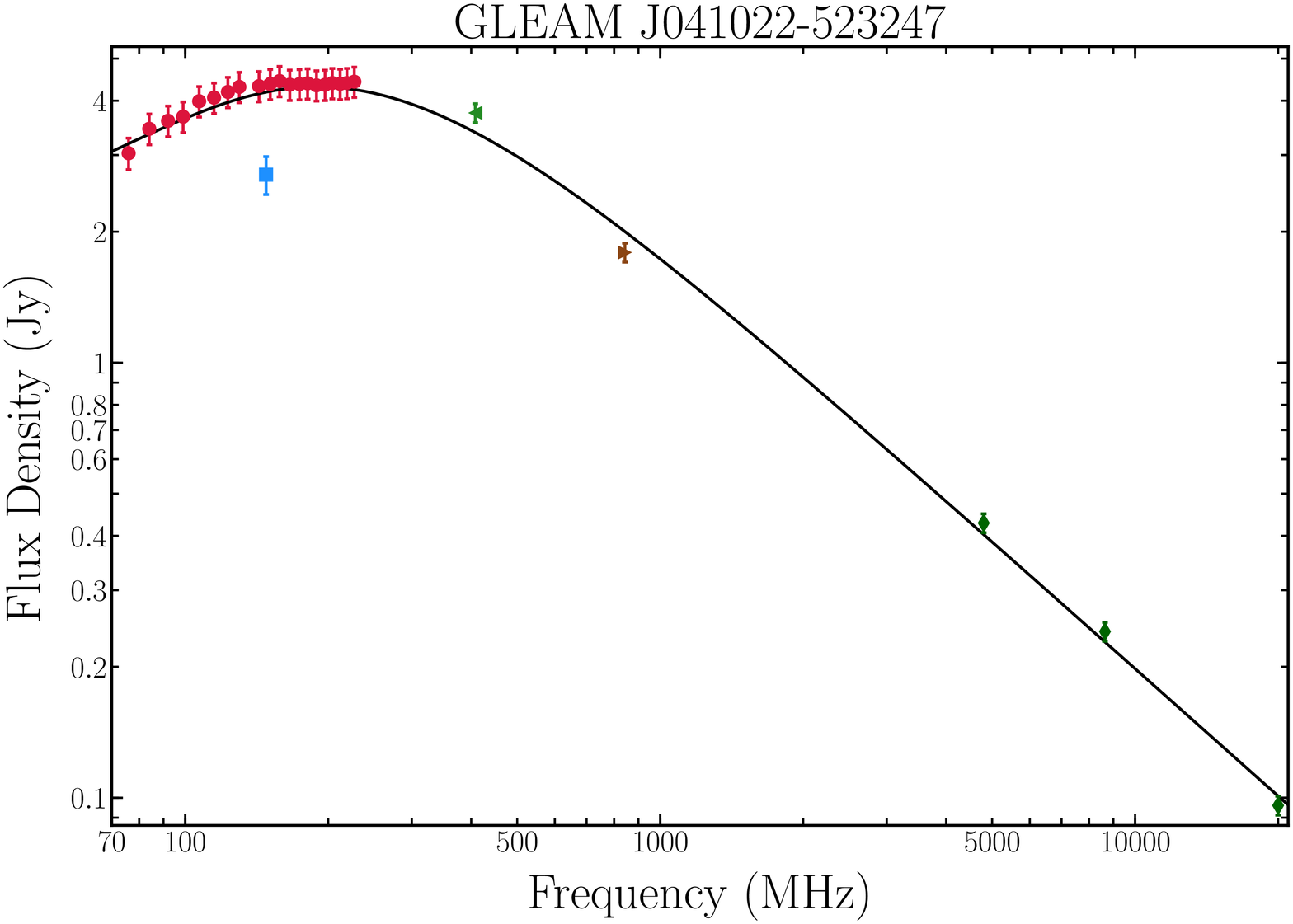}
\vspace{-0.2cm}
\caption{Light-curves and SEDs for candidate variables identified in MWATS. The upper panel of each light-curve shows the flux density as a function of sequential epoch number. The dashed grey line denotes an epoch number at which the time difference to the previous observation was greater than 8 days. The lower panel shows the flux density as a function of calender date. The red line shows the linear fit to the averaged data points. The variability statistics for the raw ($M$) and time-averaged ($\overline{M}$, $\nabla_{S}$) light-curves are given in the appropriate panel. The SEDs are shown to the right, the red data points show the 20 point low-frequency spectrum measured from the GLEAM EGC. Archival data points are also shown and described in section \ref{LC_section}.}
\label{Lightcurves}
\end{figure*}

\begin{figure*}
\centering
\includegraphics[scale=0.7]{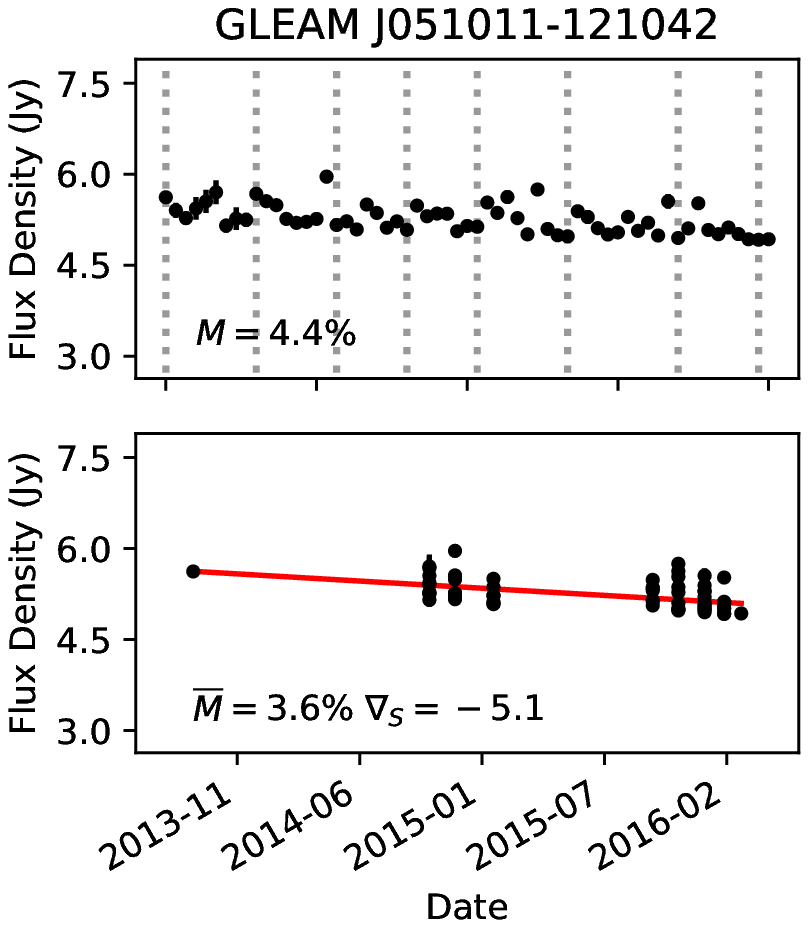}
\includegraphics[scale=0.32]{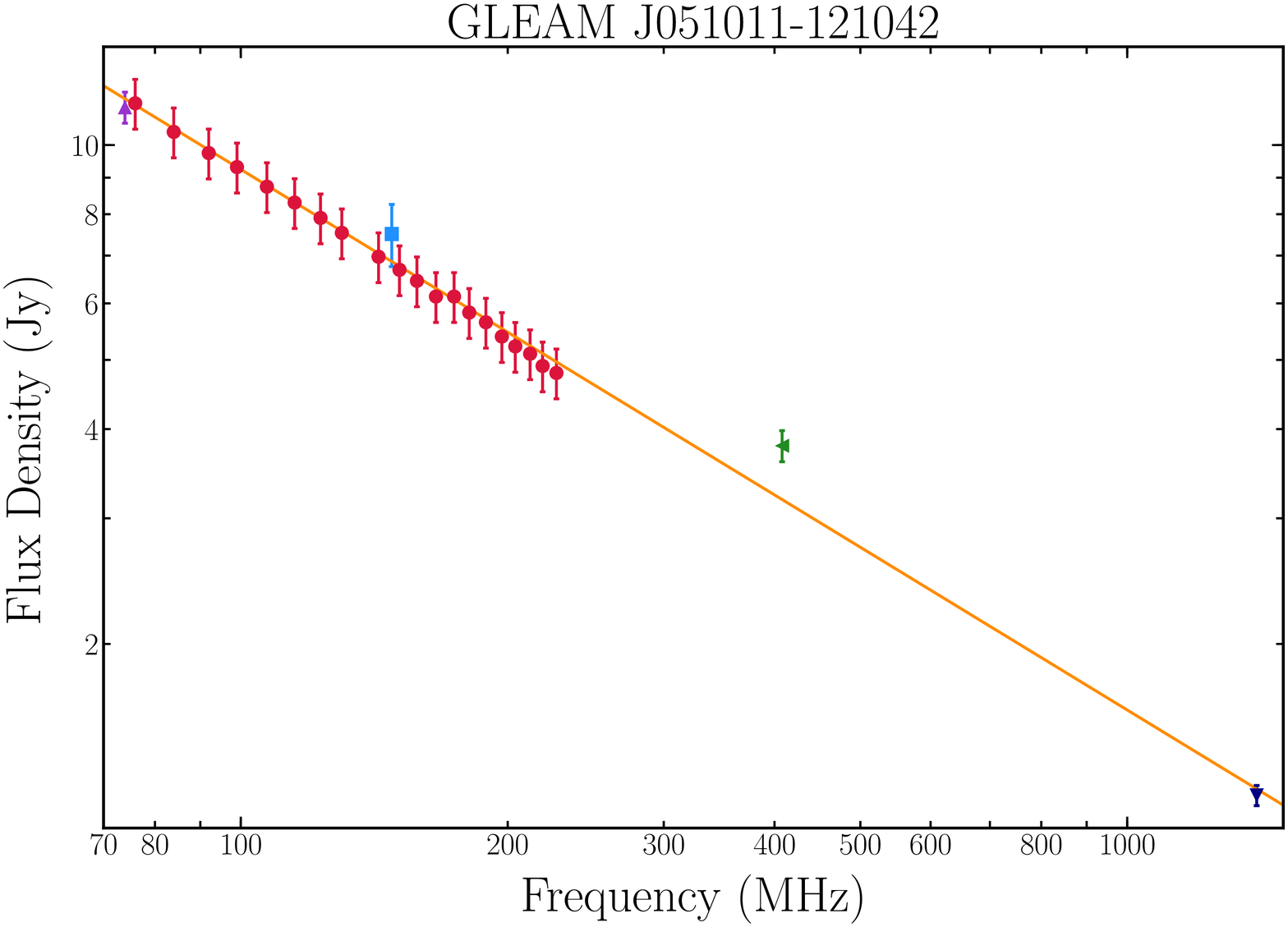}
\includegraphics[scale=0.7]{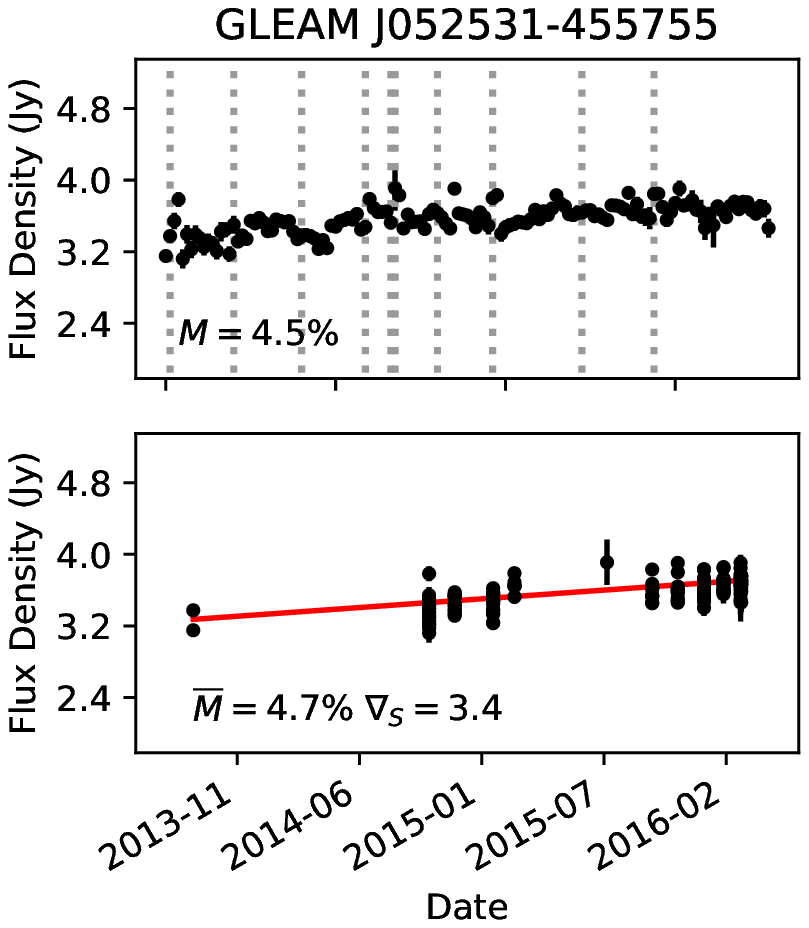}
\includegraphics[scale=0.32]{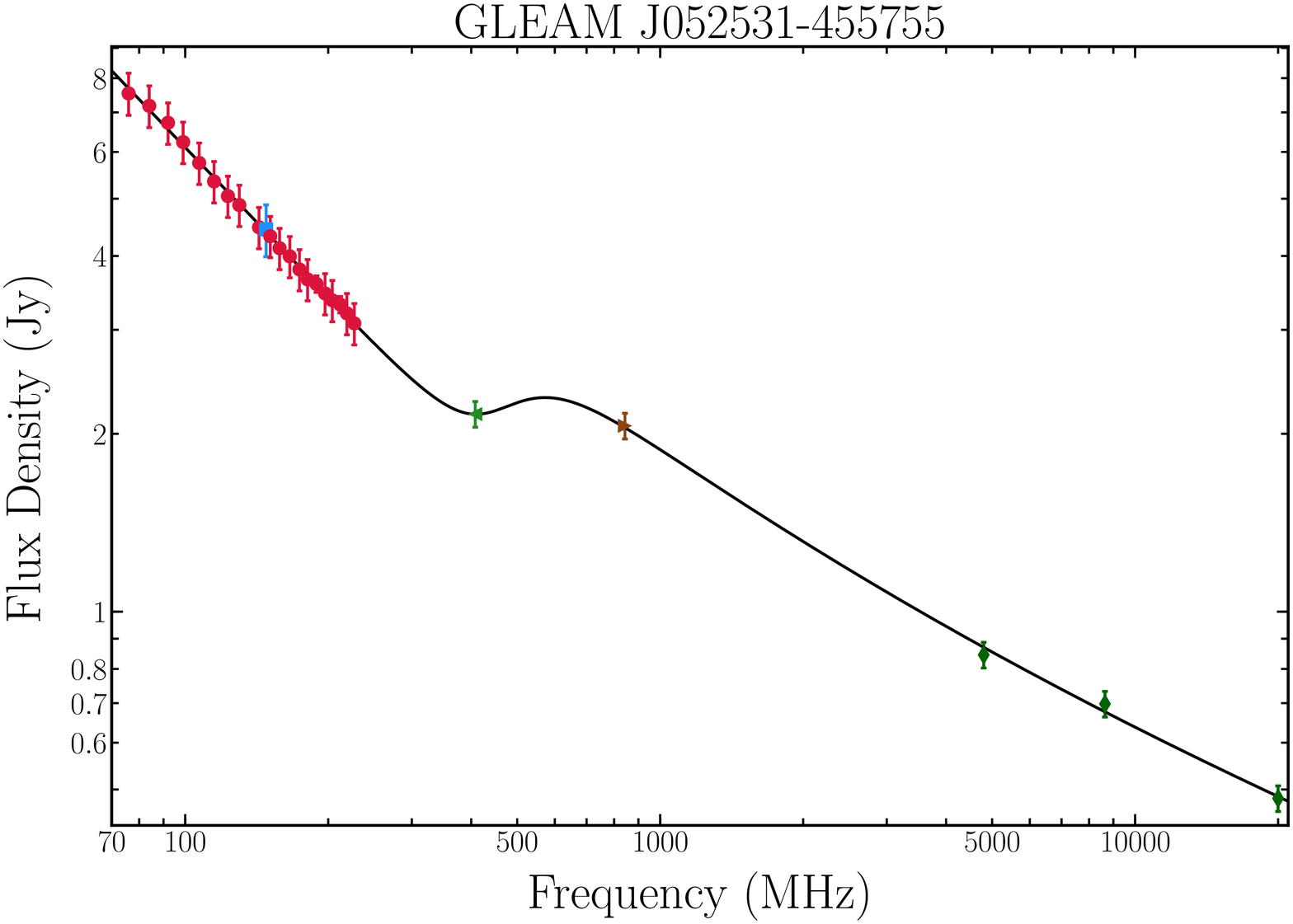}
\includegraphics[scale=0.7]{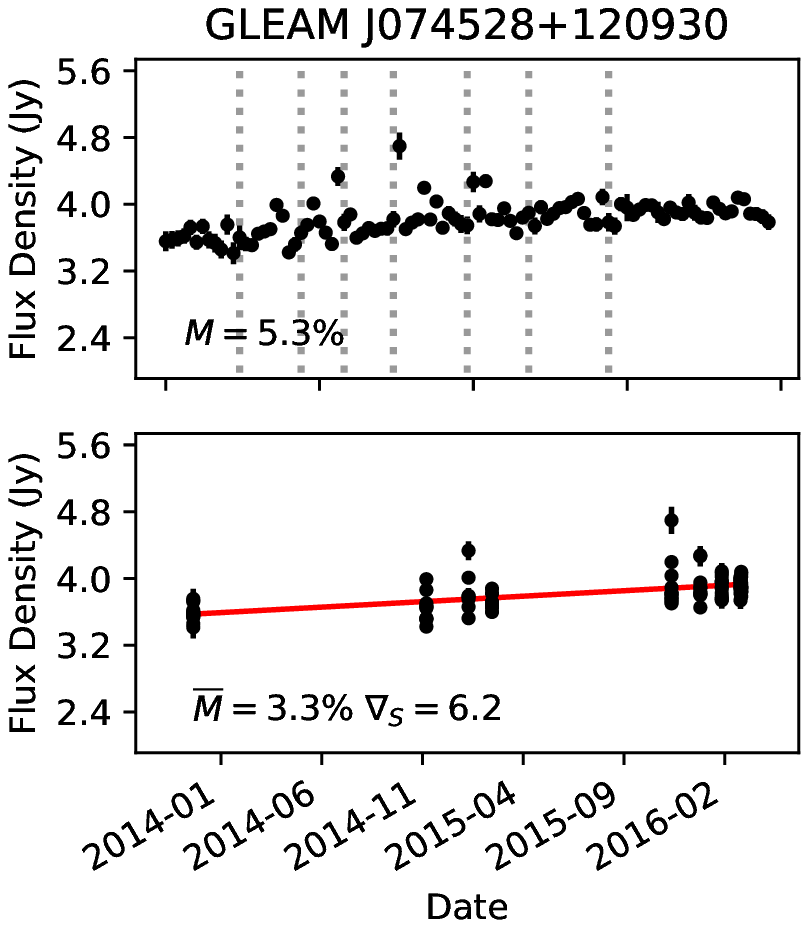}
\includegraphics[scale=0.32]{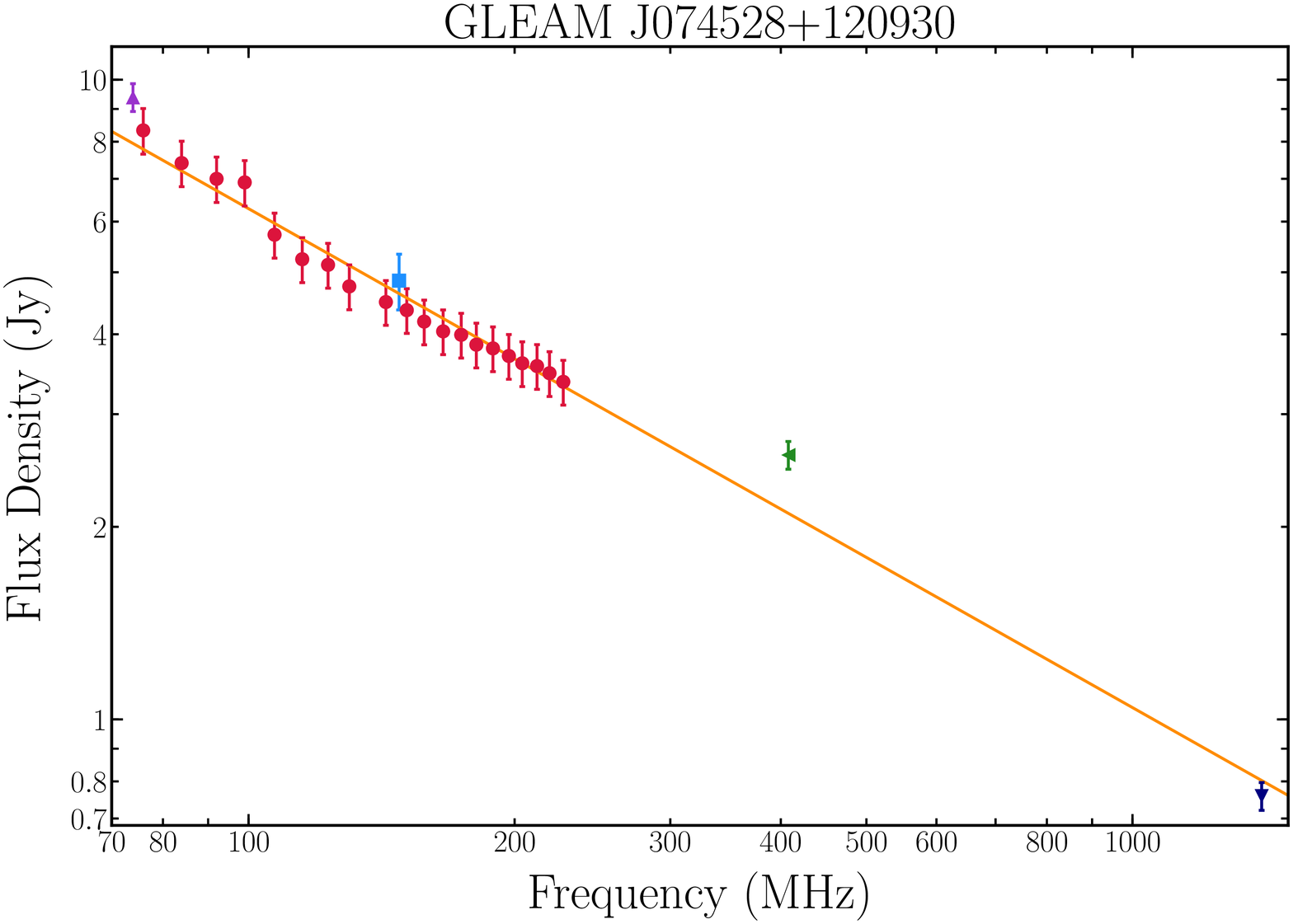}
\caption{Continued: Light-curves and SEDs for variables. Details as in Figure~\ref{Lightcurves}.}
\label{Lightcurves_2}
\end{figure*}

\begin{figure*}
\centering
\includegraphics[scale=0.7]{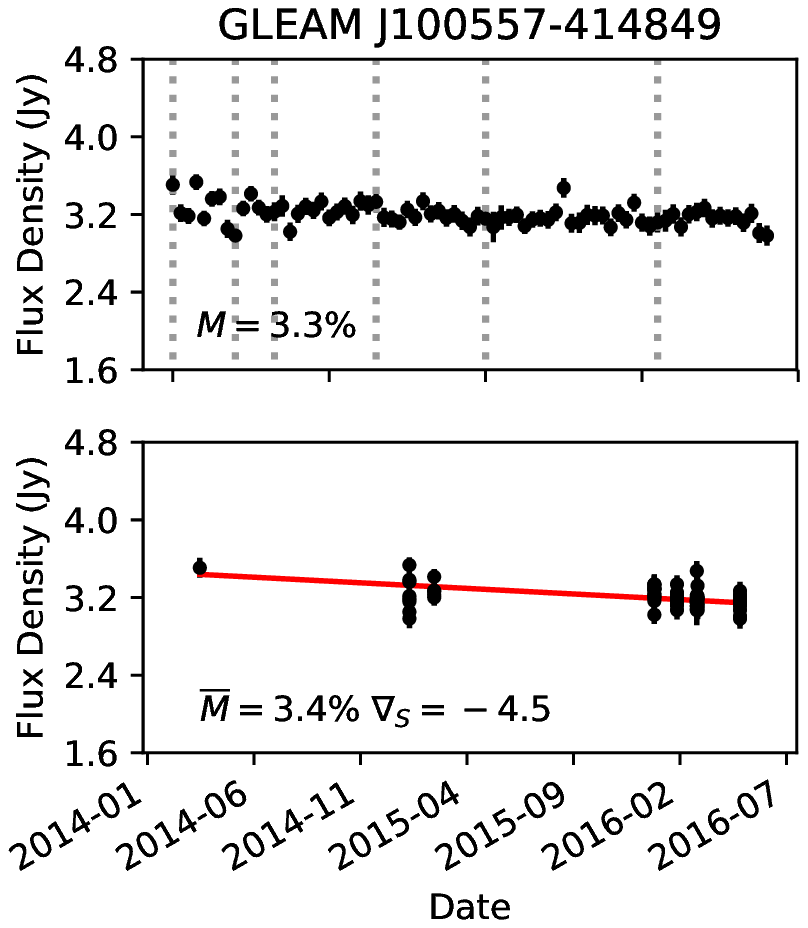}
\includegraphics[scale=0.32]{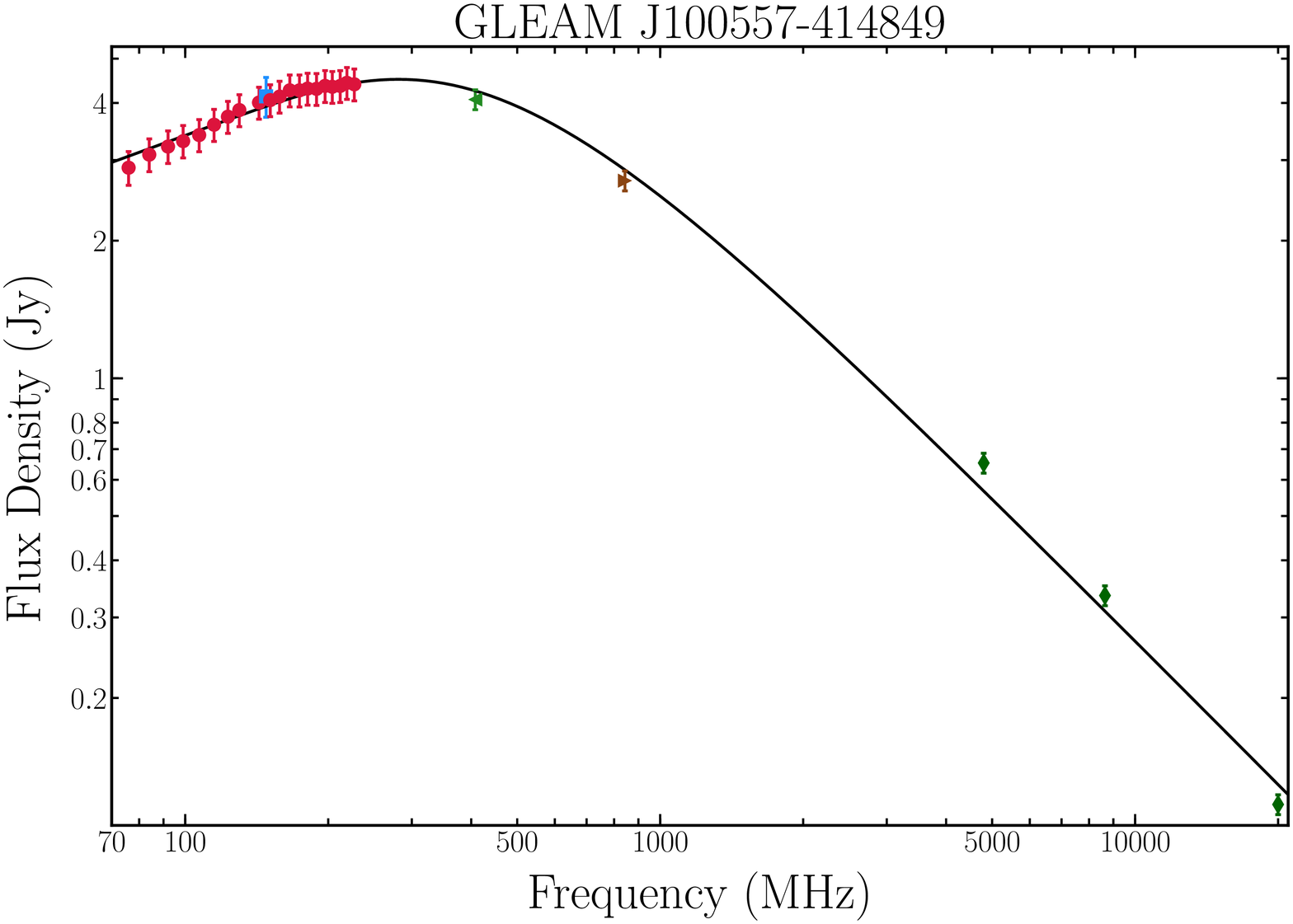}
\includegraphics[scale=0.7]{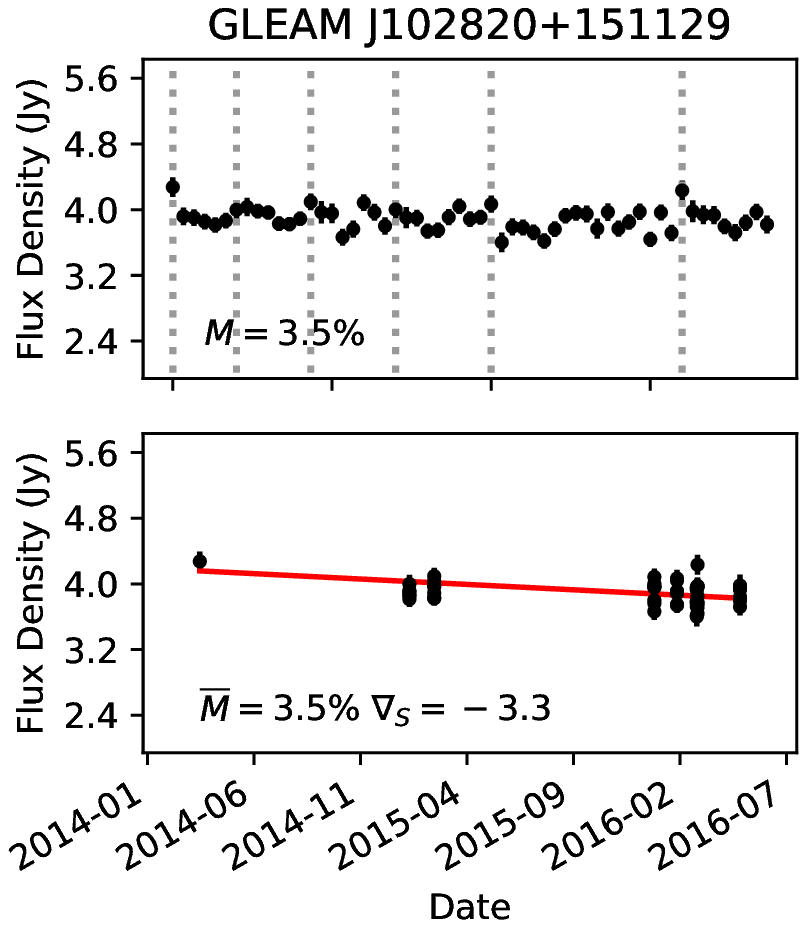}
\includegraphics[scale=0.32]{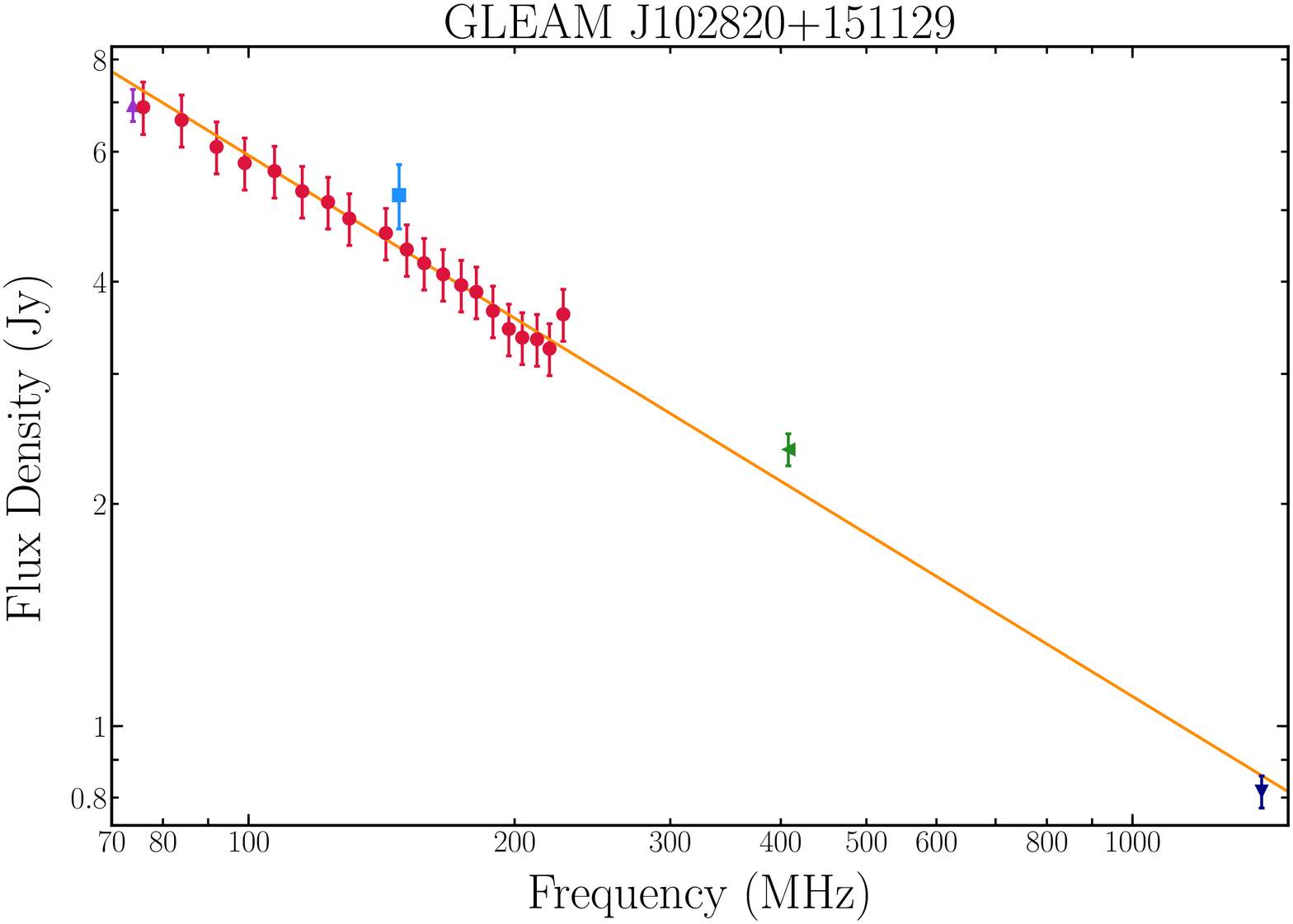}
\includegraphics[scale=0.7]{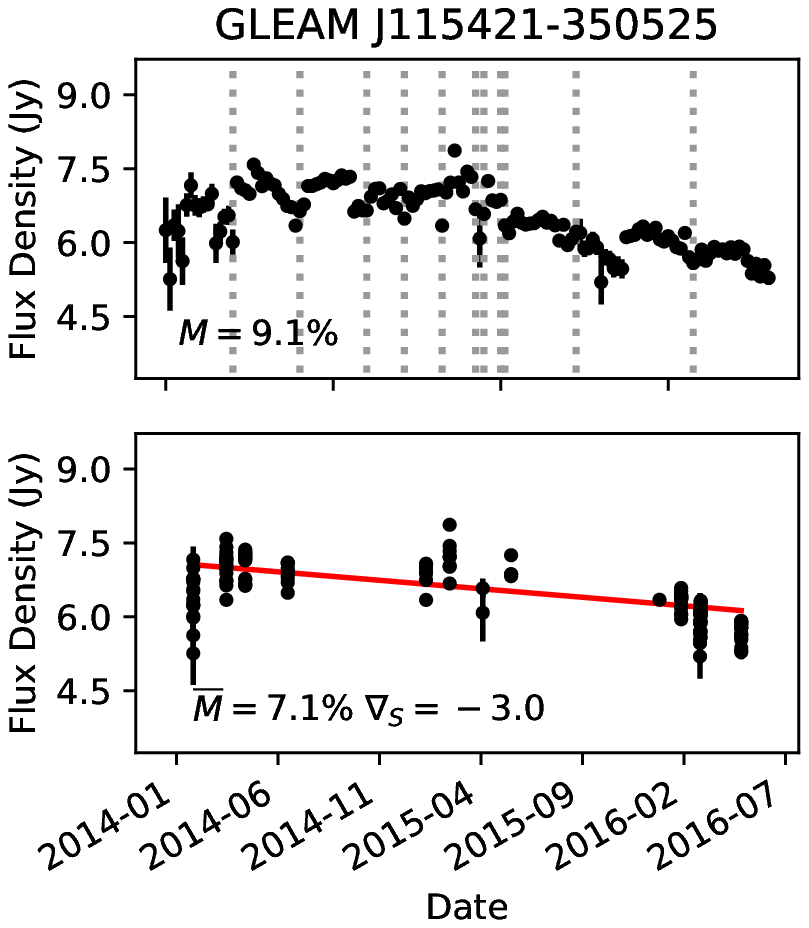}
\includegraphics[scale=0.32]{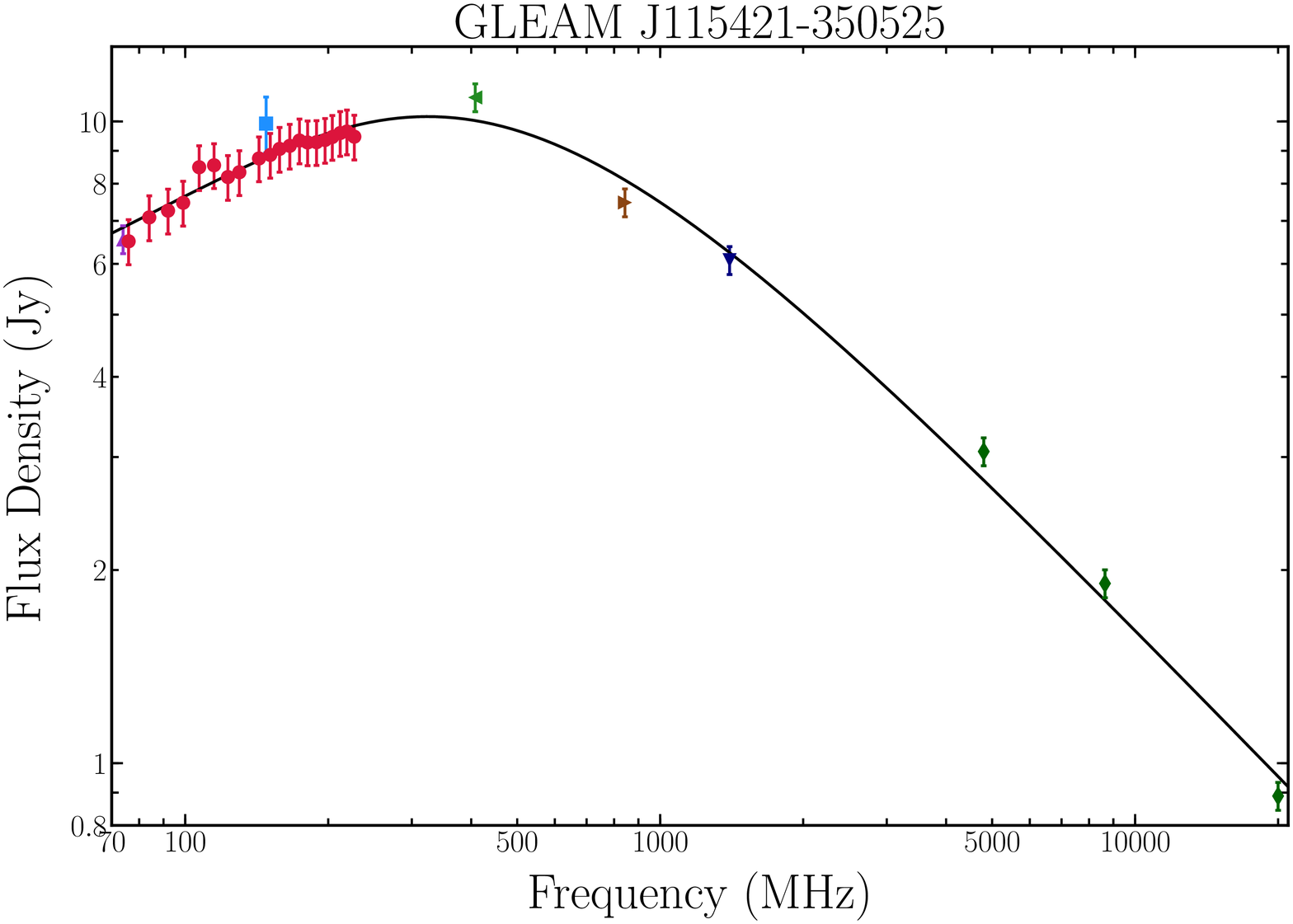}
\caption{Continued: Light-curves and SEDs for variables. Details as in Figure~\ref{Lightcurves}.}
\label{Lightcurves_3}
\end{figure*}

\begin{figure*}
\centering
\includegraphics[scale=0.7]{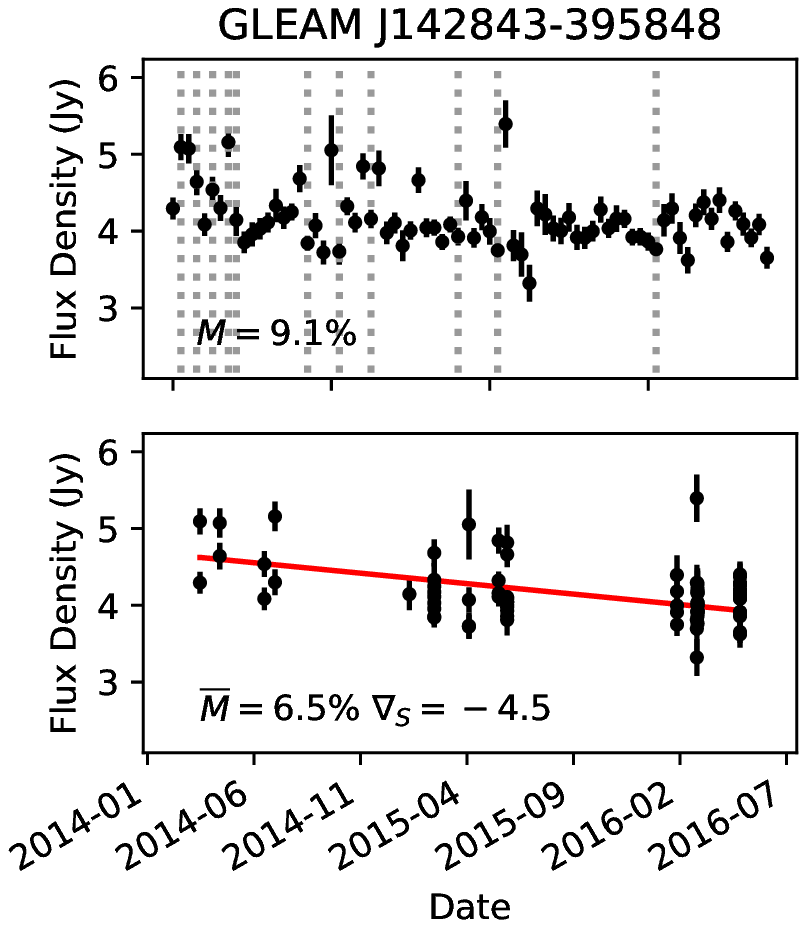}
\includegraphics[scale=0.32]{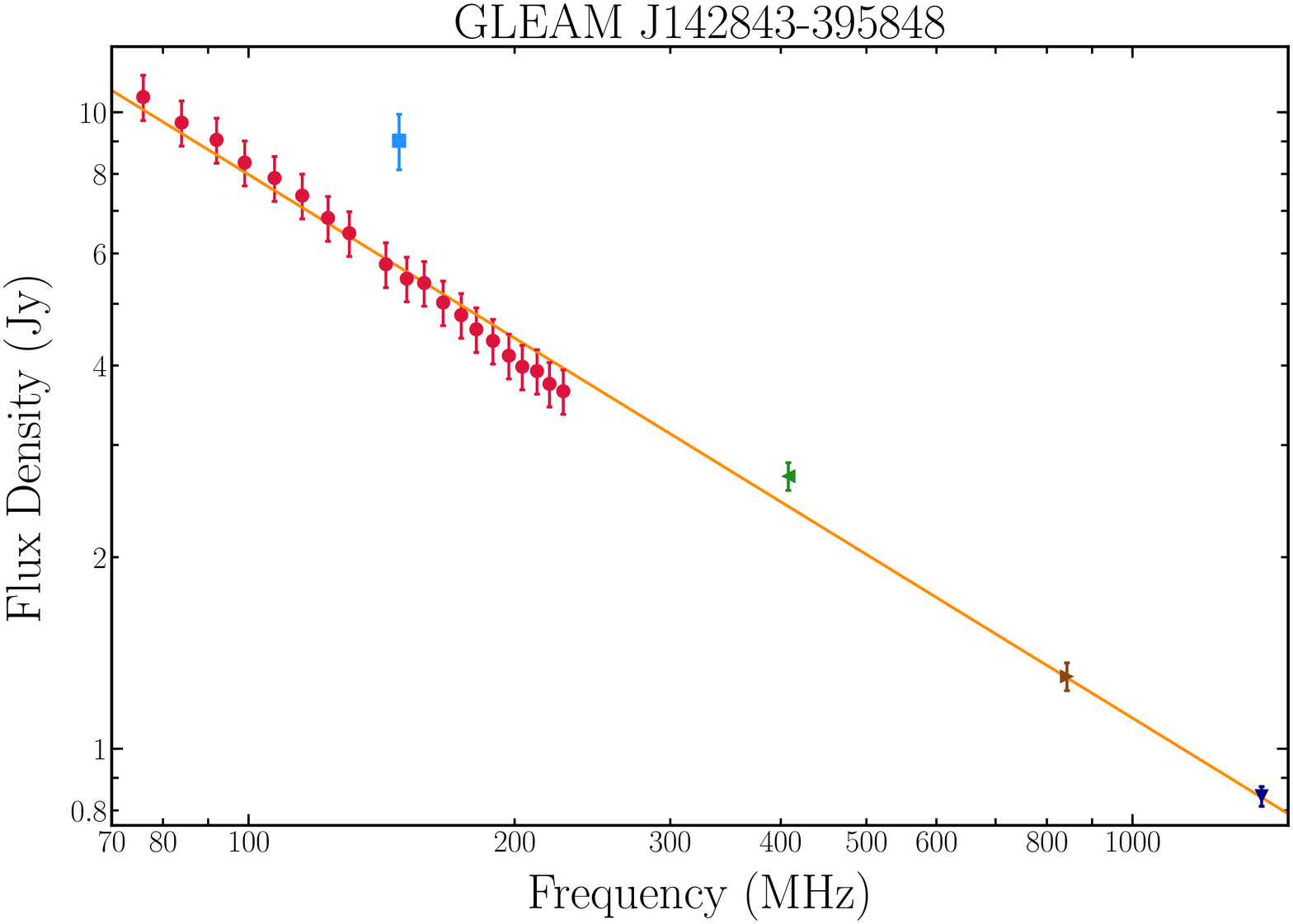}
\includegraphics[scale=0.7]{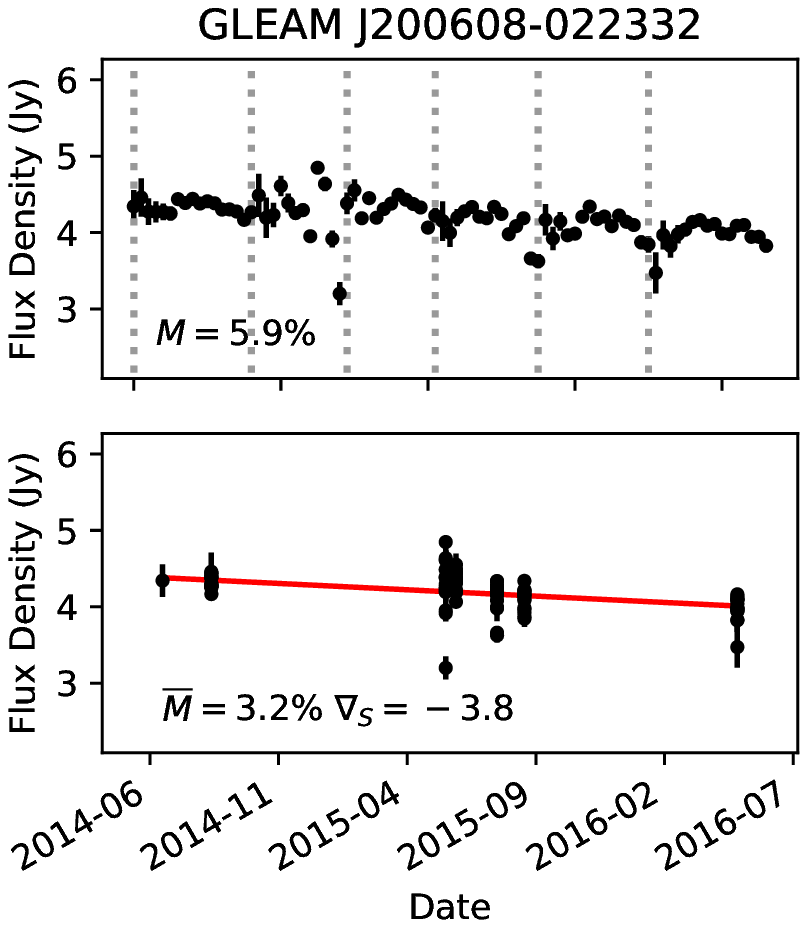}
\includegraphics[scale=0.32]{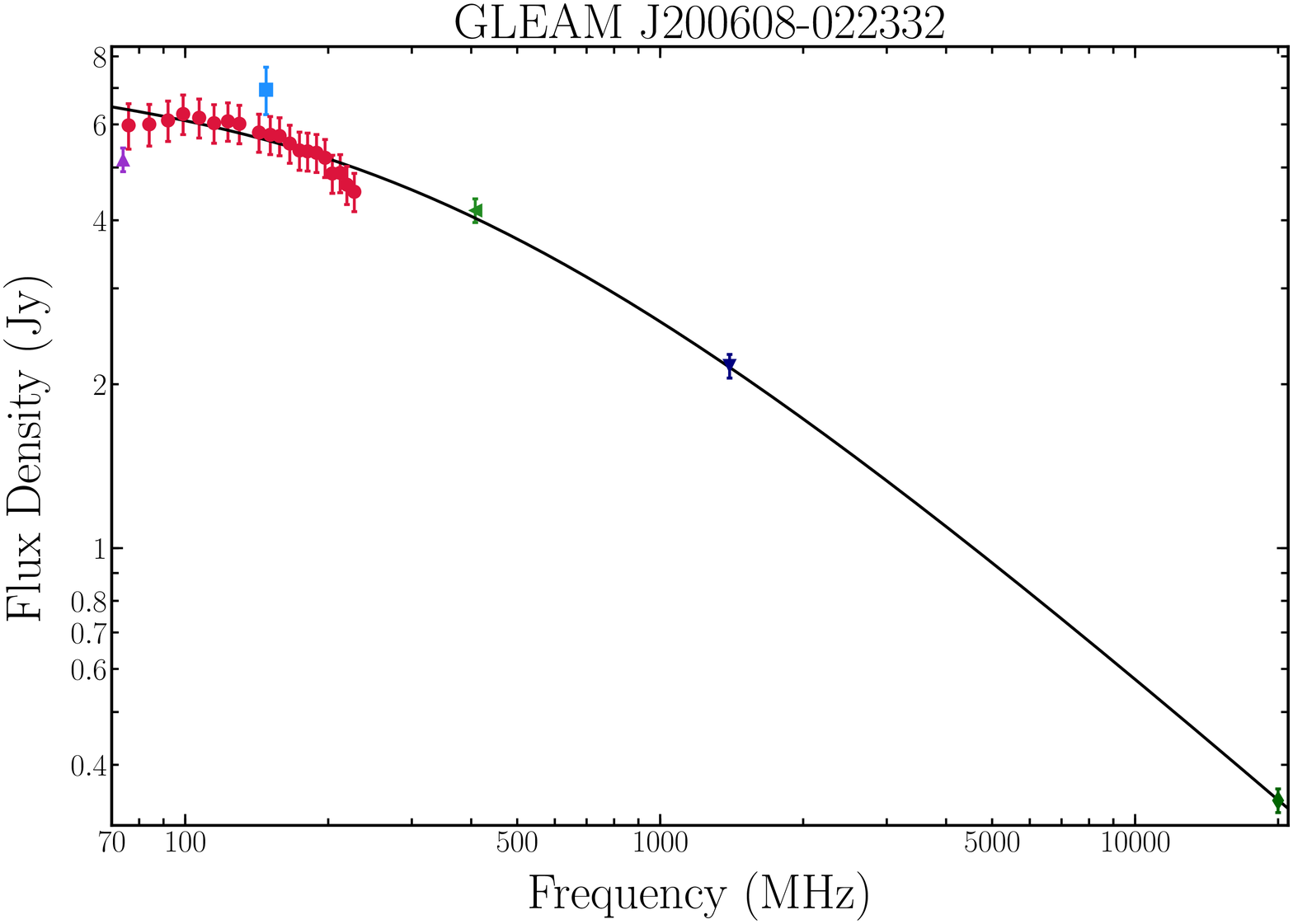}
\includegraphics[scale=0.7]{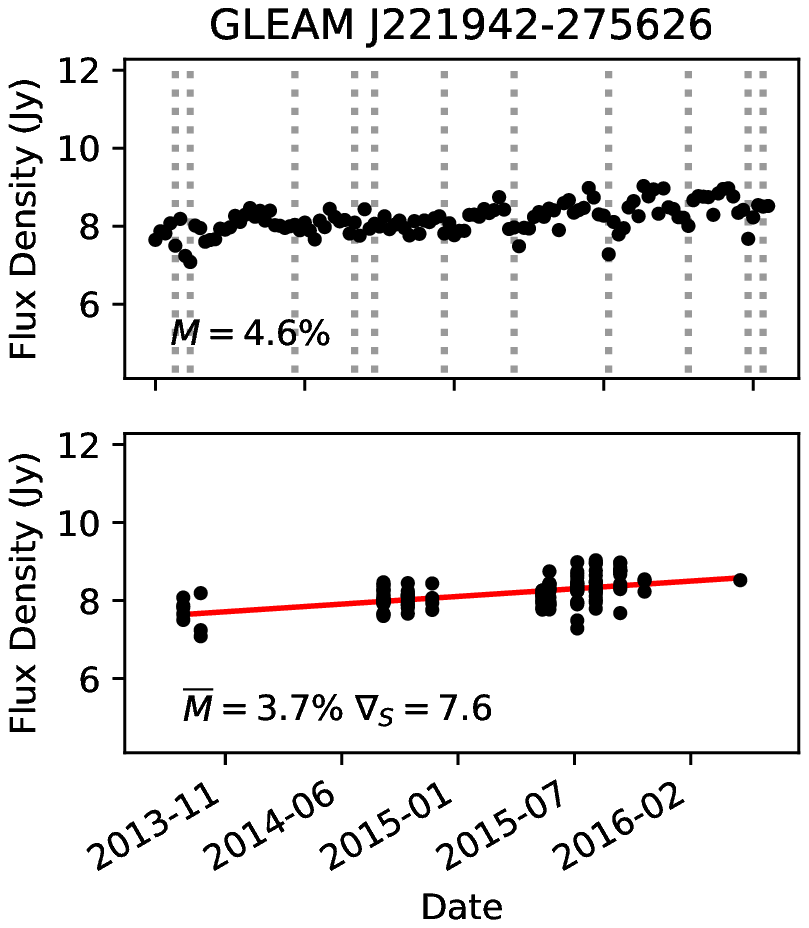}
\includegraphics[scale=0.32]{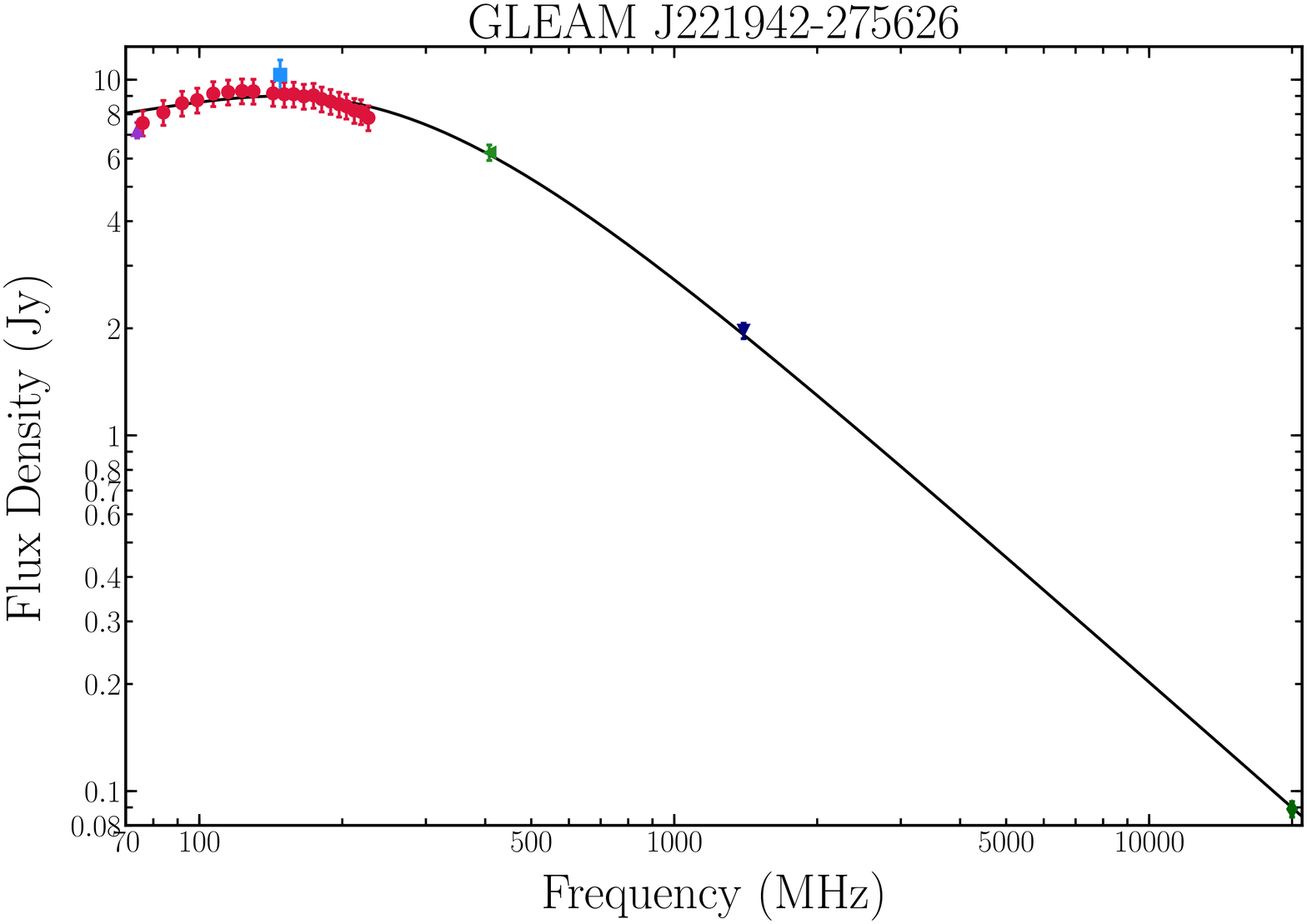}
\caption{Continued: Light-curves and SEDs for variables. Details as in Figure~\ref{Lightcurves}.}
\label{Lightcurves_4}
\end{figure*}

\begin{figure*}
\centering
\includegraphics[scale=0.7]{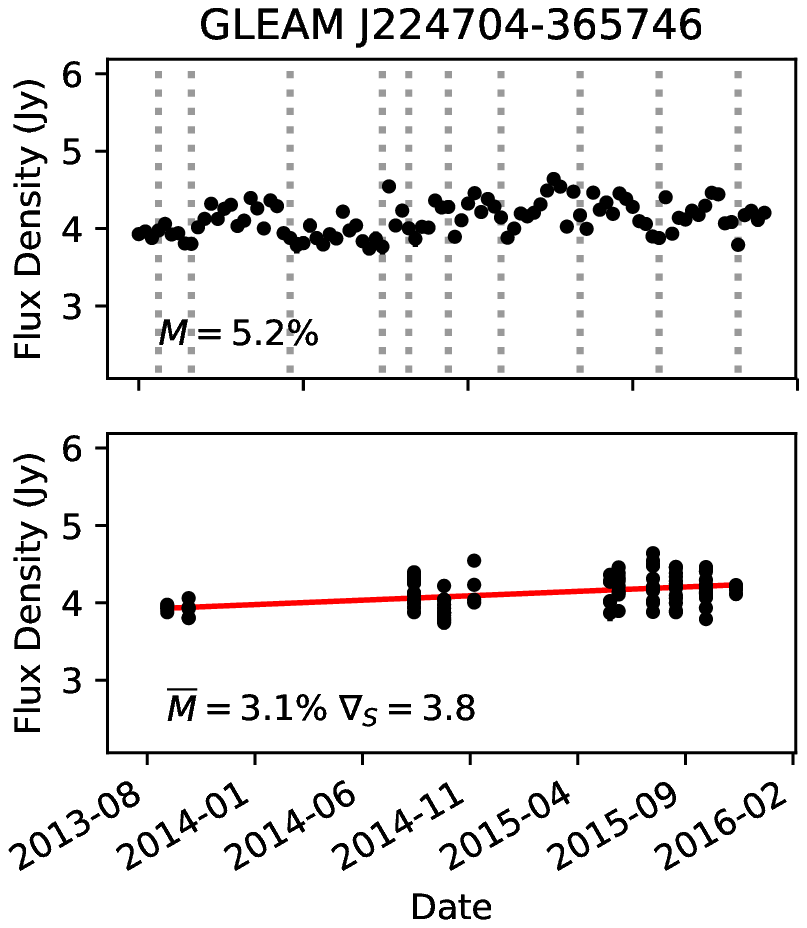}
\includegraphics[scale=0.32]{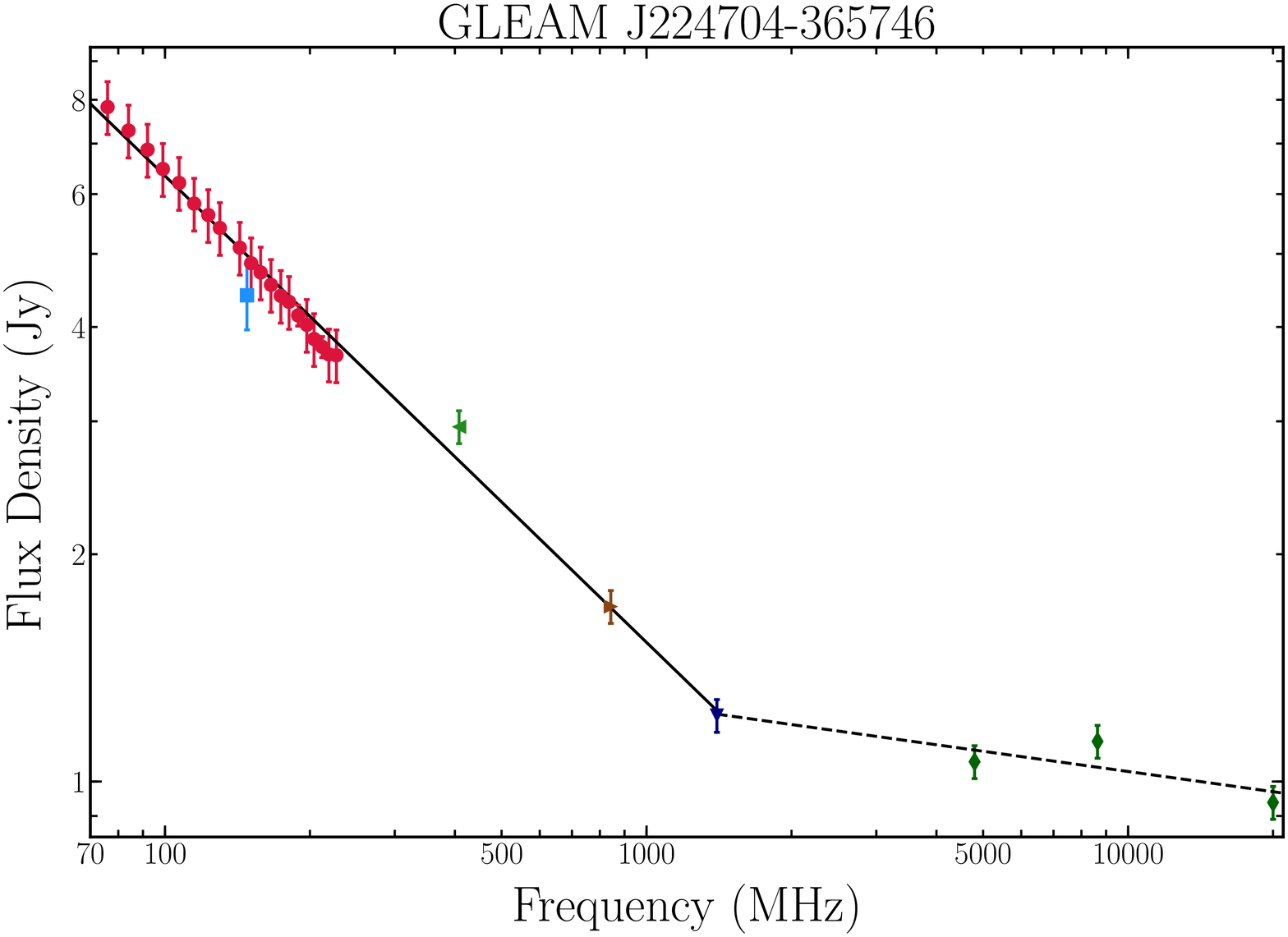}
\includegraphics[scale=0.7]{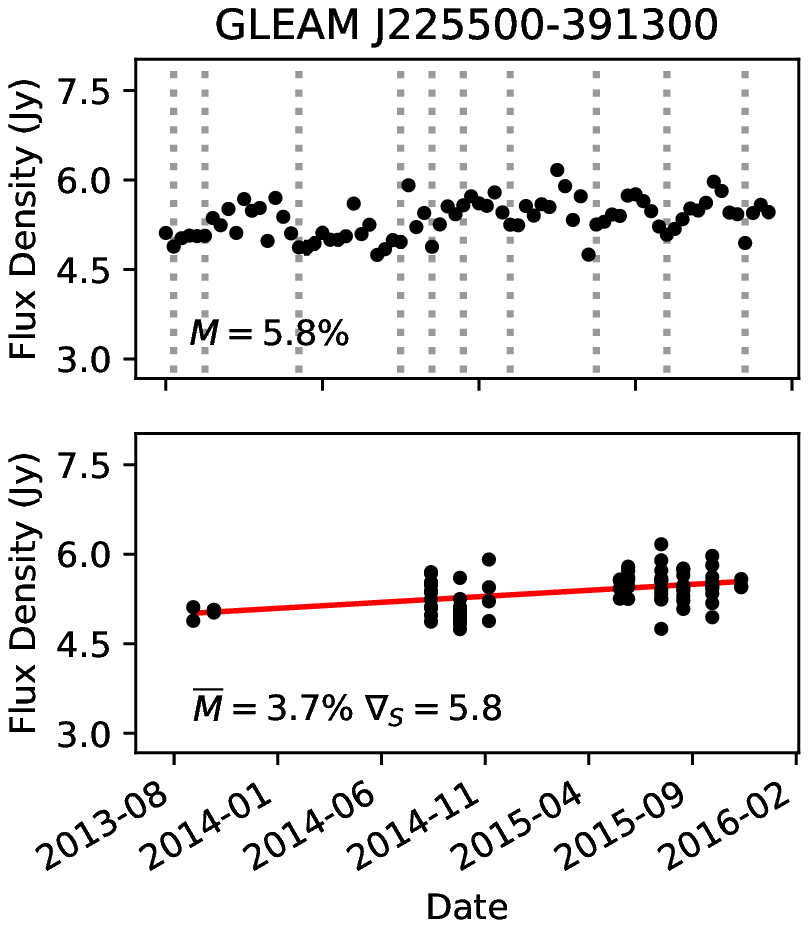}
\includegraphics[scale=0.32]{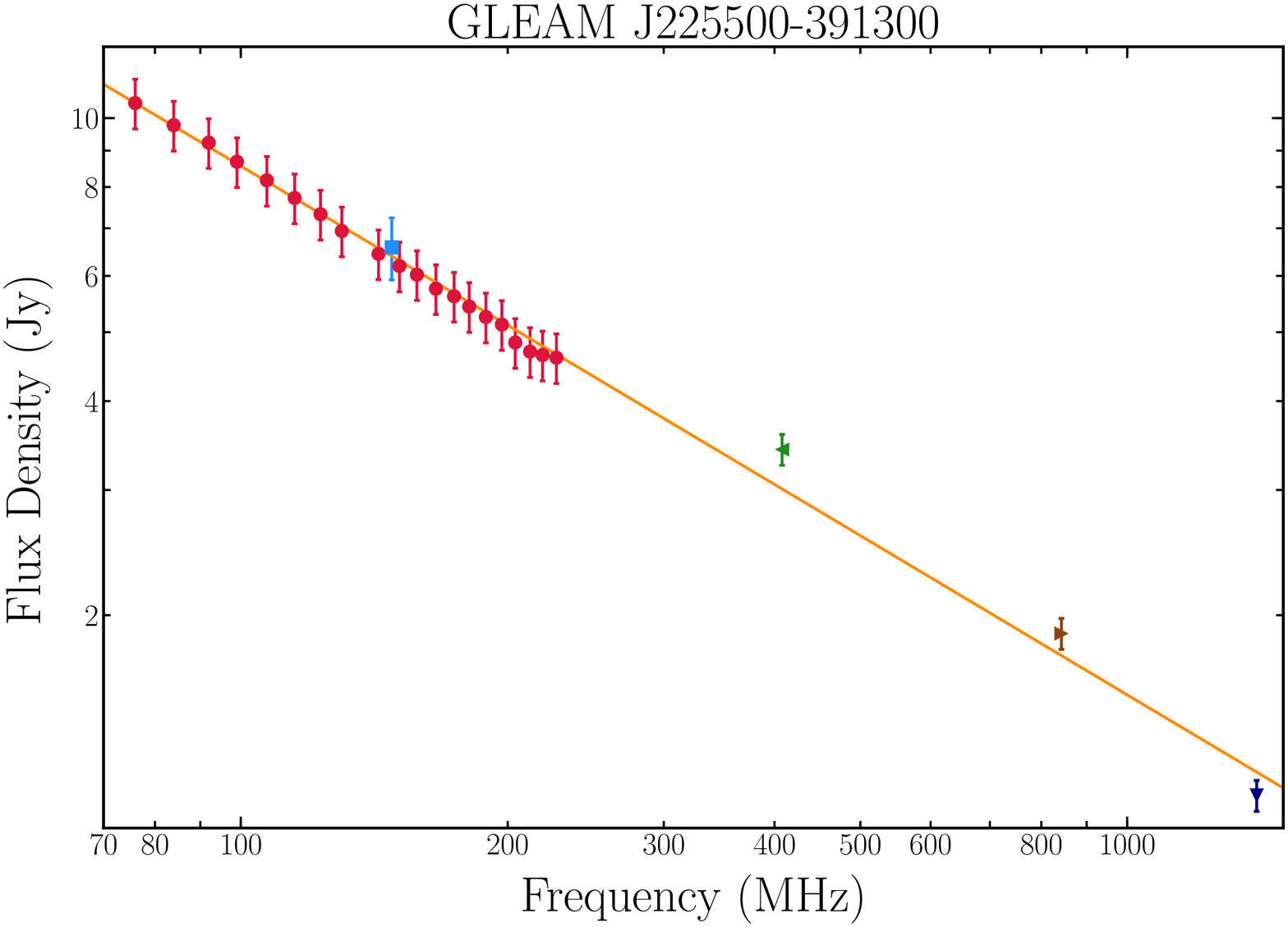}
\includegraphics[scale=0.7]{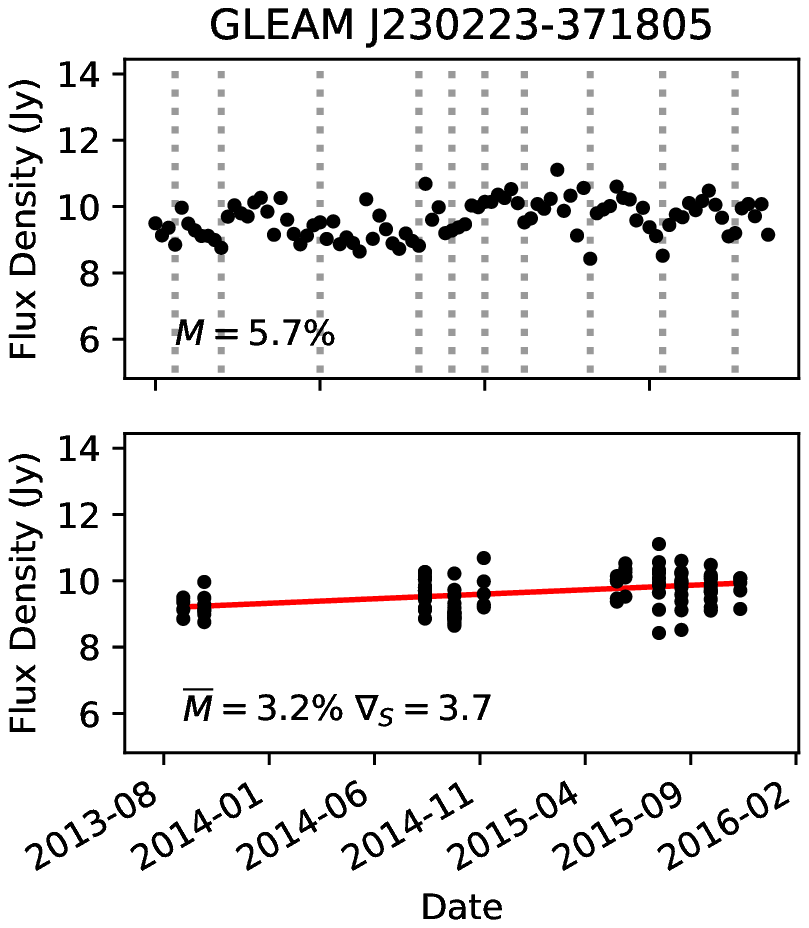}
\includegraphics[scale=0.32]{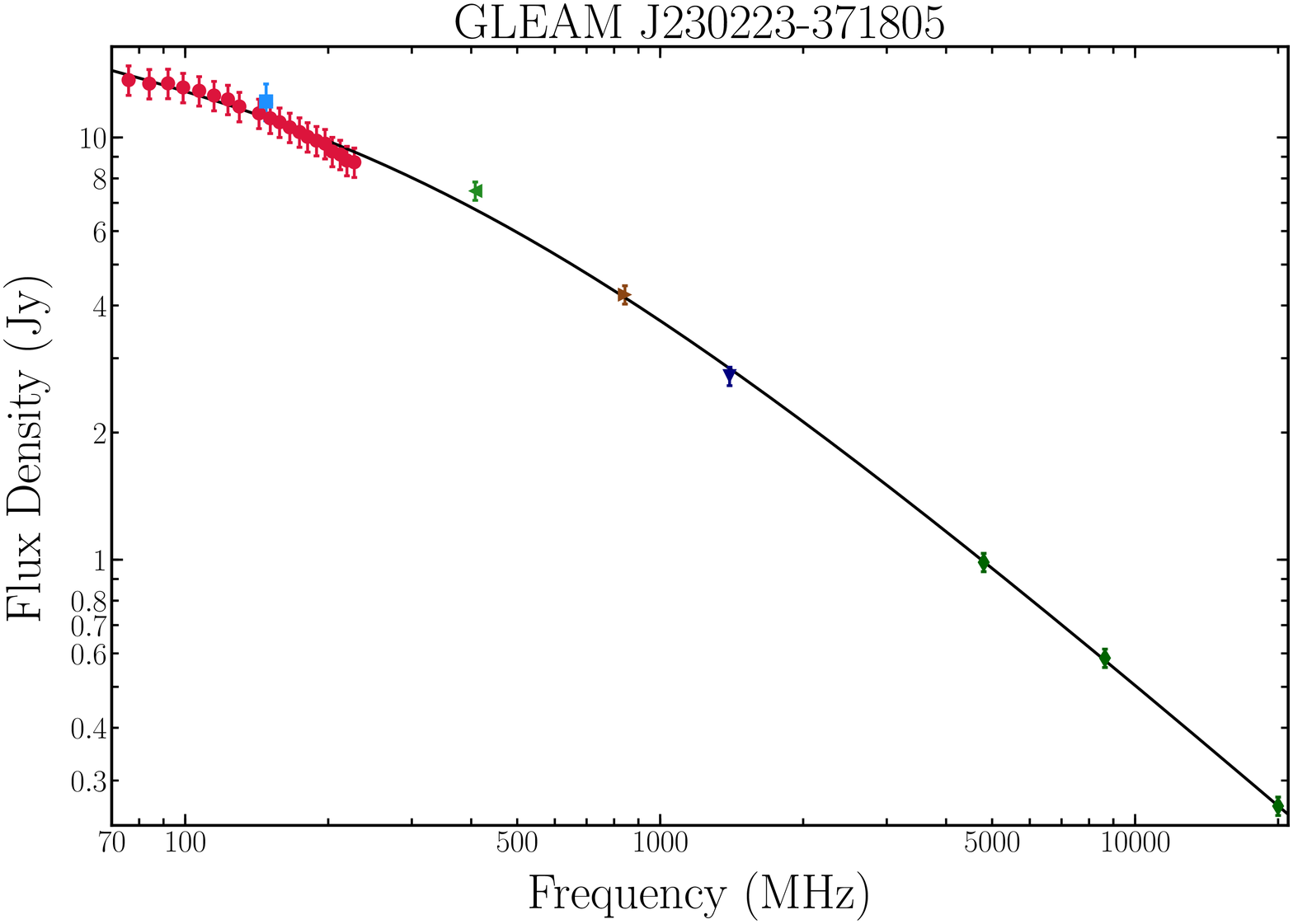}
\caption{Continued: Light-curves and SEDs for variables. Details as in Figure~\ref{Lightcurves}.}
\label{Lightcurves_5}
\end{figure*}

\subsection{Notes on specific variable sources}
\label{var_notes}
We now discuss the individual properties of the sources that showed significant low-frequency radio variability over the period covered by our study. For visual comparison, in Figure \ref{non_var} we show an example non-variable source (GLEAM J065813$-$654451) which has an averaged modulation index $\overline{M}=0.9\%$ and $\nabla_{S}=-0.1$. 


\subsubsection{GLEAM J010838+013454 (PKS\,B0106$+$013)}
This is a well studied bright Quasar/Blazar (J0108$+$0135; \citealt{Massaro_2009}) with redshift $z=2.107$ \citep{Hewett_1995} and emission from radio to $\gamma$-rays \citep{Healey_2008}. The variability detected through this survey is slow and moderate ($\overline{M}=3.1$\%). The SED is characterised by a steep power law at frequencies below 500~MHz. Examination of data in the NASA/IPAC Extragalactic Database\footnote{https://ned.ipac.caltech.edu} (NED) shows a flat SED between 1 and 100 GHz, and strong evidence for variability. 

This object is slightly too far north to have been observed by the AT20G survey, but it is included in the ICRF and VSOP surveys. Using VSOP observations \cite{Dodson_2008} give a core size of 0.2 mas at 5~GHz. 

It is detected as a variable in the MASIV survey with a modulation index of 1.9\% on time-scales of four days at 4.8~GHz \citep{Koay_2011}. In a 2.5 year monitoring campaign with the Australia Telescope Compact Array, \cite{Tingay_2003} report modulation indices of 23\%, 11\%, 4\% and 4\% at 8.6, 4.8. 2.5 and 1.4~GHz, respectively.              

\subsubsection{GLEAM J032320+053413  (PKS\,B0320$+$055)}
Best et al. (2003) identify this object as a radio galaxy at redshift $z=0.1785$. They also note that it is a compact ($<$0.2\, arcsec) radio source, with a high-ionization spectrum. A 1.6\,GHz VLBI image made by \cite{Liu_2007} in 2006 March shows a strong extended component with a weaker extended component separated by 123\,mas. \cite{Liu_2007} note that both of these components are probably lobe emission, and that their lobe separation is consistent with a Compact Symmetric Object (CSO) radio source. 

The light curve shows a gentle positive slope of variability ($\overline{M}=4.3$\%) with significance $\nabla_{S}=6.3$ over the entire time-scale of observations (see Figure~\ref{Lightcurves}). \cite{Cui_2010} carried out 5~GHz monitoring over a period of several years and found no significant variability. This object has been monitored for 15 years at 408\,MHz \citep{Bondi_408_15Years} with a modulation index of 3.2\%. Also at 408\,MHz, over a period of 5 years \cite{Fanti_81} report a fractional variability $\Delta S / S=10\%$. 

This is a well known peaked-spectrum source at low frequencies (see also \citealt{Joe_2017}). It is detected in the NVSS survey and the pronounced nature of the peak in the spectrum is quite sharp using our fitting technique. We have included in the SED the averaged data point of $6.49 \pm 0.21$ Jy reported by \cite{Bondi_408_15Years}, which is shown by an orange hexagon.  

\subsubsection{GLEAM J041022$-$523247 (PKS\,B0409$-$526)}
GLEAM J041022$-$523247 has an averaged modulation index of $\overline{M}=6.6$\% and has a peaked spectrum at low frequencies (see also \citealt{Joe_2017}). There is an AT20G counterpart but no published VLBI measurements to confirm the compact nature of the radio source. This source has a P-quad = 0.99 meaning that a 2nd order polynomial fit is preferred. We note, however, that the polynomial fit is most likely biased by a small number of data points around September 2013. The TGSS flux density measurement is significantly lower than the GLEAM EGC measurements; this could be a result of variability due to the time difference between the two surveys (TGSS was observed between 2010-2012). 

\subsubsection{GLEAM J051011$-$121042 (PKS\,B0507$-$122)}
GLEAM J051011$-$121042 shows a negative slope of variability with significance $\nabla_{S}=-5.1$. The spectral index is best described as a power law. There are no counterparts in the AT20G or ICRF catalogues, and no reported redshift in the literature but there is a faint galaxy observable in the SuperCOSMOS Sky Survey images \citep{Peacock_2016} of the field. 
The NVSS image shows an extended source (with size $35^{\prime \prime}$) and it is clearly extended in the TGSS image.  

\subsubsection{GLEAM J052531$-$455755 (PKS\,B0524$-$460)}
GLEAM J052531$-$455755 is classified as a Quasar/Blazar (\citealt{Massaro_2009}) at redshift $z=1.479$ (Stickel et al. 1993), and has been detected as a Fermi gamma-ray source (\citealt{Acero_2015}). The spectrum is best described by a steep-spectrum power-law component with a break around 400 MHz, and then continued power-law component at higher frequencies. We note our fitting technique produces a quite complex fit, which is suggestive of a spectral peak between 500 and 1000 MHz and a steep spectrum at lower frequencies. It has a counterpart in AT20G and both VSOP and ICRF catalogues. VLBI observations at 8~GHz show an unresolved compact object with size $<$10 mas \citep{Bordeaux_VLBI}. 


\subsubsection{GLEAM J074528$+$120930 (PKS\,B0742$+$122)}
GLEAM J074528$+$120930 has no published optical counterpart. The radio source shows a positive slope of variability with significance $\nabla_{S}=6.2$. The spectral index is well described by a simple power law. There is no known ICRF counterpart. 


\subsubsection{GLEAM J100557$-$414849 (PKS\,B1003$-$415) }
Burgess \& Hunstead (2006) identify this source with a galaxy at an estimated photometric redshift of $z=0.81$. It has a spectral turnover at $\sim$300~MHz and average modulation index $\overline{M}=3.4\%$. It has an AT20G detection but no published VLBI measurements. 

\subsubsection{GLEAM J102820+151129 (PKS\,B1025$+$154) }
GLEAM J102820+151129 has no published optical counterpart. The source has a VLSS measurement, no reported VLBI measurements and the spectrum is well fitted by a power-law. The MWA spectrum falls below that expected from the MRC and TGSS data points; this could potentially be a result of the difficulty in calibrating the absolute flux scales between these surveys.  

\subsubsection{GLEAM J115421$-$350525 (PKS\,B1151$-$348)  }
GLEAM J115421$-$350525 is identified with a broad-line radio galaxy at redshift $z=0.258$, and is part of the well-studied 2~Jy sample of bright radio sources (Tadhunter et al. 1993; \citealt{Morganti_1993}). The source has the highest averaged modulation index of $\overline{M}=7.1\%$ within our sample. The radio spectrum is peaked, with a turnover  at $\sim$400~MHz. There is a measurement in AT20G and it is an ICRF source. Unpublished maps\footnote{http://astrogeo.org/rfc/} show that at 2.3~GHz the core is contained within 15 mas. \cite{Bryan_2000} report variability with a modulation index of 1.9\% at 843~MHz. 

\subsubsection{GLEAM J142843$-$395848 (PKS\,B1425$-$397) }
This object has an optical identification in the SuperCOSMOS Sky Survey images \citep{Peacock_2016} but no redshift is reported in the literature. There are no AT20G or VLBI measurements. The MWA spectrum is well fitted with a power-law, but archival data points suggest a steeper power-law spectrum. This source is near ($\sim 12^{\circ}$) to the powerful radio galaxy Centaurus A, and due to its brightness it can be complex to extract measurements from this region; this could explain the discrepancy here. 
The source is strongly extended in the TGSS image and is resolved into two components in the Australia Telescope-PMN catalogue with 5~GHz flux densities 115 and 128 mJy, respectively \citep{ATPMN}. 

\subsubsection{GLEAM J200608$-$022332 (PKS\,B2003$-$025)}
This object is identified with a QSO at redshift $z=1.457$ (Aldcroft et al. 1994). There are VLSS and AT20G counterparts, but no VLBI measurements. Callingham et al. (2017) identify this as a peaked-spectrum source. It showed low levels of variability ($\overline{M}=3.2\%$) in the present study. This source is extended in the TGSS image and unresolved in  NVSS. 

\subsubsection{GLEAM J221942$-$275626 (PKS\,B2216$-$281)  }
GLEAM J221942$-$275626 is a galaxy at redshift $z=0.657$ (McCarthy et al. 1996). The source shows a steady increase in flux density with significance $\nabla_{S}=7.6$ and a low frequency turnover in the spectrum with a peak around 150~MHz. There is a counterpart in the AT20G catalogue but not in ICRF.  


\subsubsection{GLEAM J224704$-$365746 (PKS\,B2244$-$372) }
GLEAM J224704$-$365746 is a QSO at redshift $z=2.25$ (Wilkes et al. 1983). The radio spectrum is well fitted with a steep power law for frequencies below 1~GHz. At higher frequencies the data are better fitted with a flatter spectrum (shown by the dashed line). The source is detected in AT20G, and VLBA observations show that it is compact with angular diameter $<$ 0.44 mas \citep{Petrov}.  

\subsubsection{GLEAM J225500$-$391300 (PKS\,B2252$-$394) }
GLEAM J225500$-$391300 is identified with a $z=0.2615$ galaxy, observed as part of the 6dF Galaxy Survey (Jones at al. 2009). The source shows a steady positive slope in flux density with significance $\nabla_{S}=5.8$. The radio spectrum is well described by a simple power law. There is no counterpart in the ICRF catalogue but there is a counterpart in the AT20G and ATPMN catalogues. This source is slightly extended in the NVSS and TGSS images. 

\subsubsection{GLEAM J230223$-$371805 (PKS\,B2259$-$375)}
This object has a peak in the SED with a low-frequency turnover below 100~MHz (Callingham et al. 2017). It is associated with a redshift $z=1.140$ galaxy \citep{Liz_2011} and it is a point source in both the TGSS and NVSS images. It has counterparts in the AT20G and PMN surveys. It is listed as a calibrator in the VLBA database \citep{Petrov} but no information about the angular size is given.  

\begin{table*}
\begin{center}
\caption{Table summarising properties of the 15 candidate variables. The $\alpha$-Fit column gives the best spectral classification with `p-law' denoting a power-law, `Broken' denoting a two-component model and `Peak' denoting a peaked spectrum; see section \ref{var_notes} for full details. The class column gives the source class where `QSO' denotes a Quasar, `G' denotes galaxy, `BL' a Blazar and `BF' denotes a blank field where no optical identification has been made. The TGSS column gives the morphology seen in the TGSS images where, `Pt' denotes point-like, `Sl Ext' denotes slightly extended and `Ext' denotes a clearly extended source.}
\begin{tabular}{ |l|l|l|l|l|l|l|l|l|llc| } 
 \hline
 Name & $z$ & $M$ (\%) & $\overline{M}$ (\%) & $\nabla_{S}$ & P-slope & P-quad & $\alpha$-Fit & Extent & Class & TGSS \\
&  &  & &  &  &  &  & (mas) &  &  \\
 \hline
GLEAM J010838+013453 & 2.107 & 5.6 $\pm$ 0.5 & 3.1 $\pm$ 0.3 & 3.9 & 0.94 & 0.64 & Broken & $<0.2$ & QSO/BL & Sl Ext \\ 
GLEAM J032320+053413 & 0.1785 & 5.9 $\pm$ 0.7 & 4.3 $\pm$ 0.5 & 6.3 & 0.99 & 0.63 & Peak & $<200$  & G & Pt \\
GLEAM J041022$-$523247 & $-$ & 7.3 $\pm$ 0.9 & 6.6 $\pm$ 0.5 & -3.7 & 0.94 & 0.99 & Peak & $-$ & BF & Pt \\ 
GLEAM J051011$-$121042 & $-$ & 4.4 $\pm$ 0.6 & 3.6 $\pm$ 0.5 & -5.1 & 0.98 & 0.51 & p-law & $-$ & G & Ext \\
GLEAM J052531$-$455755 & 1.479 & 4.5 $\pm$ 0.7 & 4.7 $\pm$ 0.4 & 3.4 & 0.90 & 0.53 & Broken & $<10$ & QSO/BL & Pt \\
GLEAM J074528$+$120930 & $-$ & 5.3 $\pm$ 0.5 & 3.3 $\pm$ 0.3 & 6.2 & 0.99 & 0.50 & p-law & $-$ & BF & Pt \\
GLEAM J100557$-$414849 & 0.81: & 3.3 $\pm$ 0.7 & 3.4 $\pm$ 0.4 & -4.5 & 0.97 & 0.77 & Peak & $-$ & G & Pt \\
GLEAM J102820+151129 & $-$ & 3.5 $\pm$ 0.7 & 3.5 $\pm$ 0.5 & -3.3 & 0.92 & 0.95 & p-law & $-$ & BF & Pt\\
GLEAM J115421$-$350525 & 0.258 & 9.1 $\pm$ 0.8 & 7.1 $\pm$ 0.6 & -3.0 & 0.85 & 0.89 & Peak & $<15$ & G & Pt \\
GLEAM J142843$-$395848 & $-$ & 9.1 $\pm$ 1.1 & 6.5 $\pm$ 0.7 & -4.5 & 0.97 & 0.77 & p-law & $-$ & G & Ext \\
GLEAM J200608$-$022332 & 1.457 & 5.9 $\pm$ 0.5 & 3.2 $\pm$ 0.3 & -3.8 & 0.94 & 0.60 &  Peak & $-$ & QSO & Ext \\
GLEAM J221942$-$275626 & 0.657 & 4.6 $\pm$ 0.4 & 3.7 $\pm$ 0.3 & 7.6 & 0.99 & 0.50 & Peak & $-$ & G & Pt \\
GLEAM J224704$-$365746 & 2.25 & 5.2 $\pm$ 0.5 & 3.1 $\pm$ 0.3 & 3.8 & 0.92 & 0.53 & Broken & $<$0.44 & QSO & Pt \\
GLEAM J225500$-$391300 & 0.2615 & 5.8 $\pm$ 0.6 & 3.7 $\pm$ 0.4 & 5.8 & 0.98 & 0.50 & p-law & $-$ & G & Sl Ext\\
GLEAM J230223$-$371805 & 1.140 & 5.7 $\pm$ 0.6 & 3.2 $\pm$ 0.3  & 3.7&  0.92 &  0.5 & Peak & $-$ & G & Pt \\
 \hline
\end{tabular}
\label{var_table}
\end{center}
\end{table*}

\section{Discussion}
Summarising the information in Section \ref{var_notes}: of the 15 sources, five are well described by a power-law spectral index; seven have a peak and turn-over at MHz frequencies and three are not well fitted by a single spectral component. Five of the 15 sources have published VLBI measurements confirming the presence of compact components between 0.2 and 200 mas at frequencies $>2$~GHz.   
Further sources have unpublished VLBI measurements\footnote{http://astrogeo.org/vlbi/solutions/rfc\textunderscore2017d/rfc\textunderscore2017d\textunderscore cat.txt} but lack sufficient information to characterise source size. The optical properties of this sample are quite diverse: it contains four QSOs (two of which are also classed as Blazars), eight galaxies (six of which have reported redshifts), and three fields where no optical identification has been made. 

Of our total sample, 1.6\% (15/944) are classed as possible variables. There are 54 peaked-spectrum sources, which accounts for only 5.7\% of the total, yet seven sources (47\%) in our variable sample have peaked spectrum. There is therefore a significant excess in the occurrence of variability amongst peak spectrum sources. This is consistent with the findings of \cite{Rajan}, who found that every peaked spectrum source was compact on sub-arcsecond scales, and that these peaked sources dominated the compact source population. As peaked spectrum sources are often young and compact there is an increased probability for variability via refractive scintillation. 

Five objects show varying degrees of extended radio emission (see Table \ref{var_table}). One of the extended objects (GLEAM J010838+013453) has VLBI data at 5~GHz constraining the size to less than 0.2 mas. Variability in extended or resolved sources would not necessarily be predicted. However, it is difficult to completely disentangle the geometry of these objects as well as episodic periods of past radio activity. For example, GLEAM J010838+013453 could have extended radio structure from older radio activity which has expanded significantly, whilst more recent activity remains compact and unresolved. We discuss this further below with respect to the possible mechanisms for variability.    

The minimum averaged modulation $\overline{M}$ index we measure is 3.1\% and the maximum is 7.1\%. We emphasise that these sources display the largest amount of long-term variability (based on our definition) within this sample. In comparison with historical surveys, where highly variable objects are often classed as having modulation indices of $>50\%$, these objects would possibly be considered non-variable (or certainly low-level variables).  
 
 
Our selection method is biased towards detecting long-term variability, so it is possible that $\nabla_{S}$ could remain below 3.0 but short-term, high modulation index variability could still be present.  Independent inspection of high $M$ and $\overline{M}$ outliers has shown they were due to erroneous data of instrumental origin, although not every data point could be checked by hand.  We therefore cannot rule out the possibility that some of the low modulation, short-term variability could be of astrophysical origin. We can say, however, that sources with large modulation index ($M$ and $\overline{M} > 10\%$) arising from short-term variability are rare.  Future work is directed at applying more advanced statistical techniques to identify short-term variability over the entire MWATS sample.

Before we discuss the most probable mechanism of variability within this sample, it is useful to explore theoretical predictions of the time scale and amplitude expected for the cases of refractive scintillation and intrinsic jet-driven variability. We note that the debate surrounding the intrinsic versus extrinsic nature of variability in low frequency sources has largely been concluded (see \citealt{Rickett_86}). It is, however, pertinent to frame the results in this paper not only with respect to previous work but also in reference to upcoming surveys with next generation instruments such as the SKA. 

\subsection{Intrinsic variability} 
If we assume that the variability observed in our sample is  driven solely by the synchrotron process and that the Compton catastrophe limit holds \citep{Kell1969}, then the maximum brightness temperature we can achieve is $T_{b}<10^{12}\,$K (assuming there is no beaming). We can therefore define the following expression (\citealt{MJ_2008}): 

\begin{equation}
T_{b} = \dfrac{S_{\nu} c^{2}}{2k_{B} \nu^{2} \Omega} \leq 10^{12}\,{\rm K},
\label{Tb_eq}
\end{equation}

\noindent where $S_{\nu}$ is the amplitude of variability, $k_{B}$ is the Boltzmann constant, $\nu$ is the observing frequency and $\Omega$ is the solid angle subtended by the source. The solid angle can be approximated by $\Omega = r^{2} / D^{2}$, where $r$ is the source size and $D$ is the distance. We assume that $r$ and thus the angle subtended by the source is limited by the light travel time (or time-scale for variability) $\tau$, i.e., $r=\tau c$. By substituting in $\Omega = D^{2}/\tau^{2}c^{2}$ we can then define the following expression, which gives the maximum flux density change permitted for a synchrotron emitting source: 

\begin{equation}
\Delta S \leq \dfrac{2k_{B} \nu^{2}\tau^{2} T_{b}}{D^{2}}.
\end{equation}

\noindent If we take $\tau = 2.8$\,years (i.e.\ the maximum time-scale sampled in this survey) and $D=10$\,Gpc (a reasonable assumption given our sample contains objects with known distances between $\sim1-15$ Gpc) we find $\Delta S < 5.3 \times 10^{-4}$~Jy.

\begin{figure*}
\centering
\includegraphics[scale=0.7]{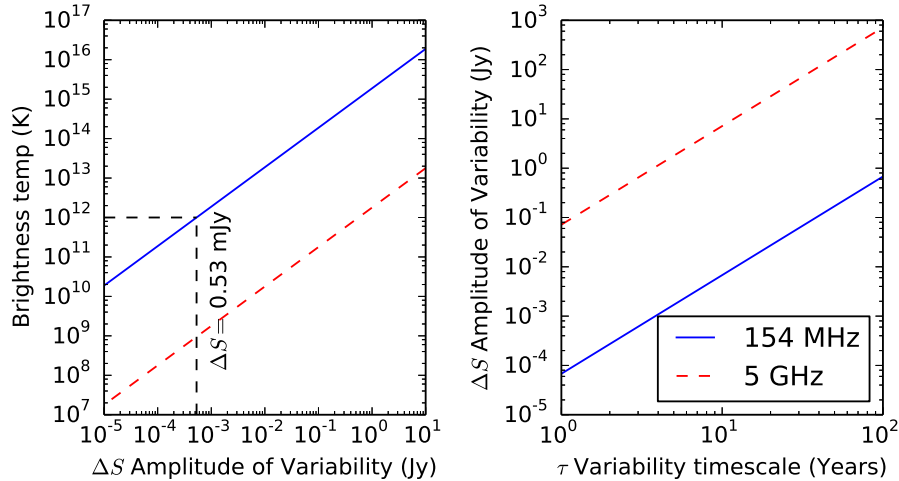}
\caption{Left panel: A plot of amplitude of variability ($\Delta S$) versus brightness temperature for a hypothetical source. We assume in this example that the source is at a distance of 10~Gpc and the variability time-scale is 2.8 years. To not break the brightness temperature limit of $10^{12}K$, the amplitude of variability ($\Delta S$)  is restricted to be below 0.53~mJy (at 154~MHz). Right panel: A plot of variability time-scale ($\tau$) versus amplitude of variability ($\Delta S$). In this case we assume that $T_{b}=10^{12}$K and we find that for typical variations of $\Delta S=0.5$~Jy we require variability time-scales to be greater than $\sim$100 years (at 154~MHz). The comparison case of 5~GHz is shown with the red dashed line).}
\label{BT}
\end{figure*}

In reality we measure changes that are much greater i.e., $\Delta S \sim 0.5$\,Jy; see the left panel of Figure \ref{BT} for a plot of the expected brightness temperature as a function of amplitude of variability ($\Delta S$). Assuming $\Delta S \sim 0.5$\,Jy and recalculating the permitted time-scales of variability we find $\tau> 100$\,years. See the right panel of Figure~\ref{BT} for a plot of variability time-scale $\tau$ as a function of amplitude of variability ($\Delta S$). To obtain brightness temperatures below $10^{12}$\,K the amplitude of variability is therefore quickly quenched and varies as $T_{b} \propto \tau^{-2}$. 

Relativistic beaming and Doppler boosting can increase the maximum brightness temperature permitted in these scenarios and scales as $b^{3}T_{b}$ (see \citealt{Boosting}). Assuming a boosting factor of $b=10$ and re-calculating the example above, we find allowed flux density variations to be as large as $\Delta S = 0.53$~Jy, with time-scales of $\tau>3.16$\, years. This is much more in-line with our results. Two of our variables are classed as Blazars and it is plausible that some level of beaming is taking place and that this is contributing to the changes in flux density. We note that these objects are also compact and have angular diameters less than 0.2 and 10 mas respectively (see Table~\ref{var_table}). In the next section we will discuss how the compact nature of these objects can give rise to variations via refractive scintillation.   

Assuming the absence of beaming, then the calculations above provide a strict limit on the permitted variability time-scales and flux density changes at low frequencies. As relativistic beaming is rare, it is unlikely that simple synchrotron radiation can explain the type of variability seen in our survey, as the variability time-scales are very long (as restricted by the Compton catastrophe limit) and are not well sampled by our survey.  

\subsection{Refractive interstellar scintillation}
The time-scale for refractive scintillation is a function of wavelength and scales as $\nu^{\frac{-\beta}{\beta-2} }$, where the spectral index $\beta$ is 11/3 for Kolmogorov turbulence \citep{ARS_1995}. Assuming at 5\,GHz we observe scintillation time-scales as fast as one day, at 154\,MHz we would expect these time-scales to be 5.8\,years. The modulation index of variability scales with frequency as $(\nu / \nu_{0})^{17/30}$, where $\nu_{0}$ is the transition frequency and the exponent arises from assuming Kolmogorov turbulence \citep{Walker_1998}.  

The transition frequency separates weak and strong scattering, with weak scattering dominating the higher frequencies. The transition frequency is a function of Galactic latitude and is as high as 40\,GHz within the Galactic plane and as low as $\sim 6$\,GHz around the Galactic poles (\citealt{Walker_1998}, \citealt{Walker_2001}). Assuming a modulation index of 100\% at the transition frequency, at 154\,MHz we would predict a modulation index of 12.7\% at the Galactic pole (with $\nu_{0}=6$~GHz) and 4.4\% towards the Galactic centre (with $\nu_{0}=40$~GHz). These values are consistent with our results (i.e. Median $\overline{M}=2.1\pm2.0$\%).


At low frequencies the radio spectra of compact sources can often include 
extended steep-spectrum relic emission from earlier periods of source 
activity. When combined with generally larger beam sizes, the contribution 
from compact flat-spectrum components, which may be prominent at higher 
frequencies, will therefore be strongly diluted. These factors will play 
an important role in the prevalence and detectability of low frequency 
radio variability as a function of time, frequency and angular resolution.
For refractive scintillation, above a certain angular size cut-off angular broadening will take place and the scintillation and modulation index will be quenched. The quenching amount depends on the ratio of the angular size cut-off ($\theta_{lim}$) and source size ($\theta_{s}$) as $(\theta_{lim} / \theta_{S})^{7/6}$  \citep{Walker_1998}.

If we take the object GLEAM~J010838$+$013454 as an example, VLBI observations show a core size of 0.2\,mas at 5~GHz. Using the NE2001 electron density model\footnote{https://www.nrl.navy.mil/rsd/RORF/ne2001/} \citep{NE2001} we can calculate the angular size cut-off (or angular broadening parameter) as a function of location on the sky. We find that for GLEAM~J010838$+$016453 we require the source size to be less than, or the majority of the flux density to be contained within, 48.5\,mas at 154~Mhz (calculated with a transition frequency of 8.0\,GHz). As VLBI measurements of the core show a size less than 0.2\,mas, it clearly has a compact high-frequency core. This source is, however, slightly extended in the 150~MHz TGSS image with a deconvolved major axis of $\approx 15$\,arcsec. 

The transition from a flat spectrum to a steep spectrum at $\nu \approx 400$\,MHz is shown clearly in the SED in Figure 5. It is complicated to completely untangle the geometry here, especially as a function of time and frequency. For example, relic low frequency radio emission may be present as well as more recent compact emission. In all cases, however, providing a significant fraction of the flux is contained within the angular size cut-off, refractive scintillation is still possible.  

One final possibility not fully explored here is that of variability caused by a free-free absorbing screen of material moving across the line of sight. This has been shown to cause significant spectral variability in the peaked spectrum source PKS 1718$-$649 and would possibly alleviate the need for all flux to be contained within a compact region (see \citealt{Tingay_2015} for a full discussion). 

Both the time-scale and expected modulation for refractive scintillation are in line with our results and is the plausible explanation for variability within our sample, although we cannot completely rule out a contribution from beamed synchrotron emission. In both the synchrotron (non-beamed) and refractive scintillation cases the time-scale of variability is typically longer than our survey length and justifies why fitting the gradient of the light curves is a reasonable strategy for detecting and characterising variability.      

\begin{figure}
\centering
\vspace{-0.5cm}
\includegraphics[scale=0.78]{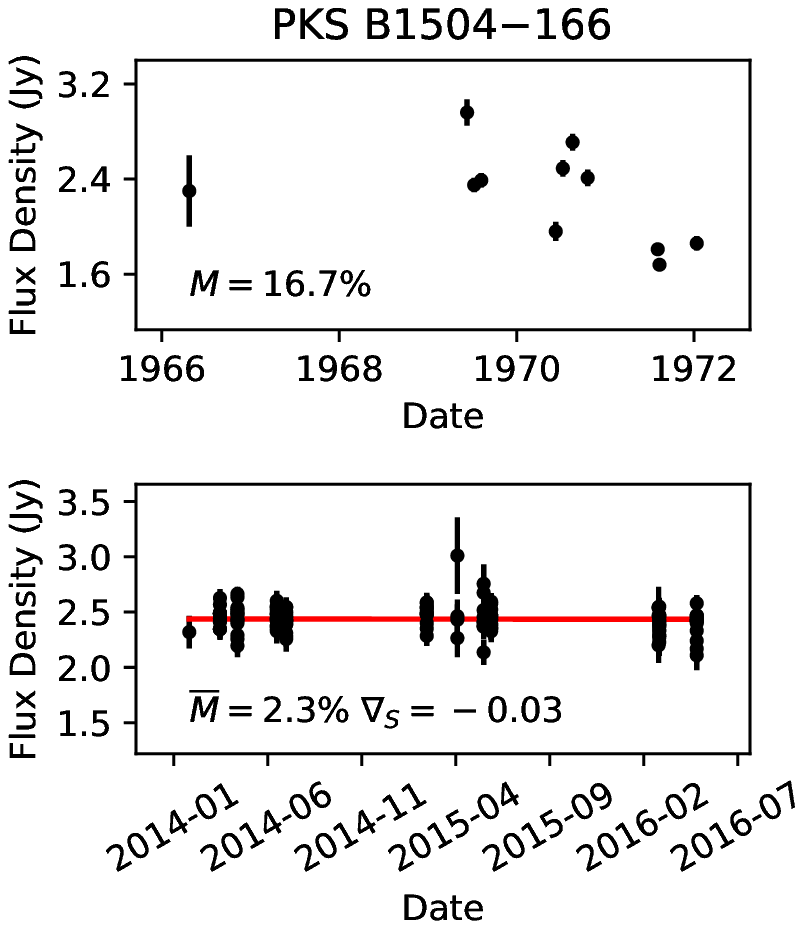}
\includegraphics[scale=0.75]{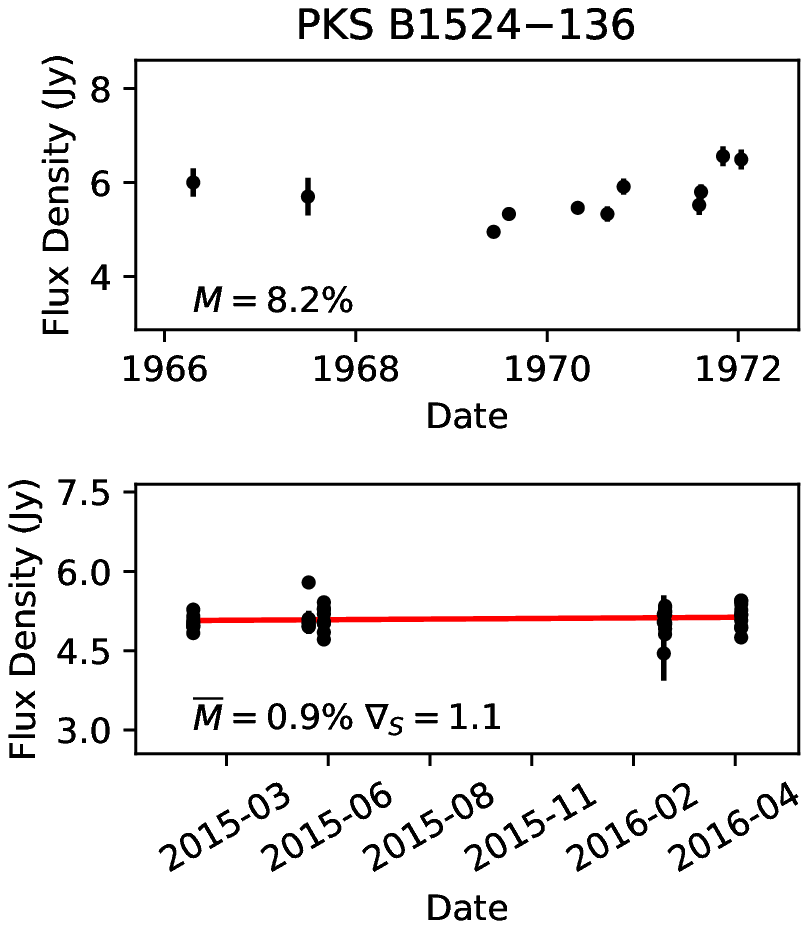}
\includegraphics[scale=0.72]{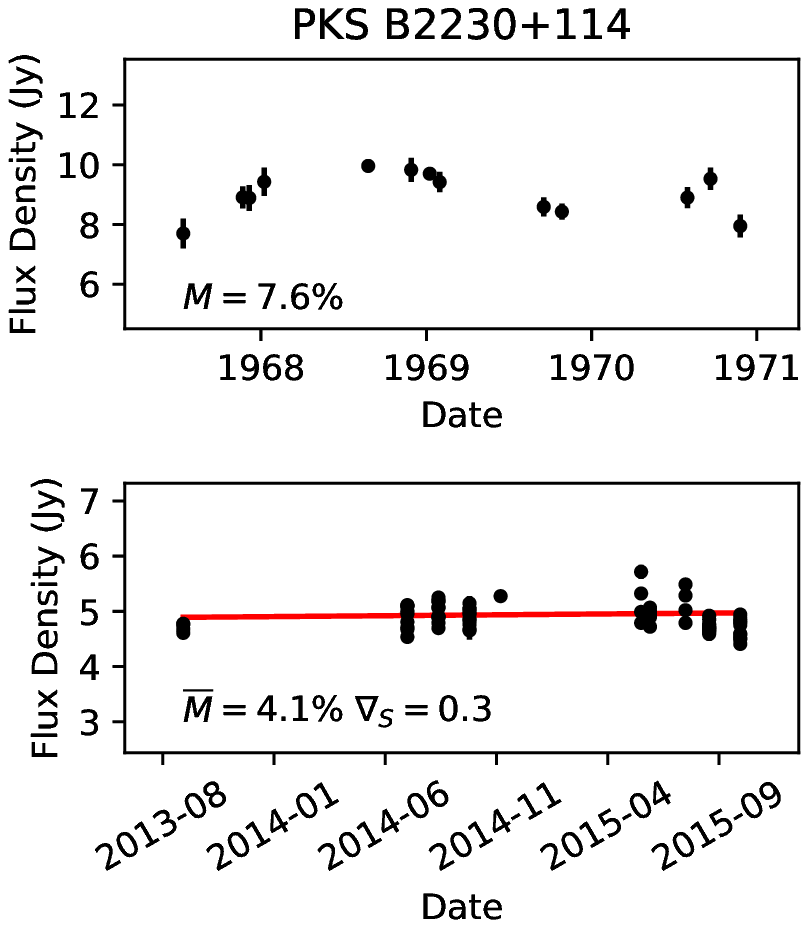}
\caption{Upper panel: Light-curves of PKS B1504$-$166, PKS B1524$-$136, PKS B2230$+$114; measurements reported in Hunstead (1972) at 408~MHz. Lower panel: MWA measurements from this survey.}
\label{Comp_lightcurves}
\end{figure}

\subsection{Comparison with previous surveys}
\label{comp}
The original \cite{Hunstead_72} paper reported four variables: PKS~B1504$-$166, PKS~B1524$-$136, PKS~B2230$+$114 and PKS~B2251$+$158 at 408~MHz. The source PKS~B2251$+$158 is located at the edge of our survey range and we did not have many measurements for this object. The remaining three sources are plotted in Figure \ref{Comp_lightcurves}. All three objects are classified as non-variables using the criteria we define in this paper. 

The source PKS~B1504$-$166 has the highest modulation index in archival observations ($M=16.7\%$) whereas we measured only $\overline{M}=2.3\%$. VLBI observations of PKS~B1524$-$136 at 8~GHz by \cite{Mantovani_VLBI} show a double sided jet with angular extent up to 300 milli-arcsec. Observations by \cite{Jeyakumar_2000} show that 50\% of the total flux density is from a compact structure with a size of $<100$ milli-arcsec. It has been monitored at 408 MHz for 15 years by \cite{Bondi_408_15Years} who report a modulation index of 8.2\%. 

The source PKS~B2230$+$114 has also been studied at 408 MHz for 15 years by \cite{Bondi_408_15Years}. We only measure an $\overline{M} =4.1$\%, whereas \cite{Hunstead_72} measured 7.6\% and \cite{Bondi_408_15Years} report a value of 4.3\%. VLBI observations by \cite{Xu_1998} constrain the core size to $<2$ milli-arcsec. The angular size limit at 154 MHz in this direction is 48 mas and the transition frequency is 9.31 GHz (calculated using NE2001; \citealt{NE2001}). Using the transition frequency and assuming 100\% modulation at this transition we would predict modulation indices of 9.8\% and 17.0\% at 154 and 408 MHz respectively. In both the 154 and 408 MHz cases here we measure lower variability than predicted, but the ratios of our measurements and those of \cite{Hunstead_72} are consistent.    

Clearly the variability of these three objects is being quenched at lower frequencies. Both PKS B2230$+$114 and PKS~B1524$-$136 have peaked spectra, whilst PKS~B1504$-$166 has a flatter spectrum, all of which are indicators of compactness. As discussed above, the amount of low frequency variability is largely a function of frequency and source size. Potentially the object PKS~B1524$-$136 has become extended such that refractive scintillation has become quenched at 154~MHz, or, that the frequency shift has reduced the modulation and lengthened the time-scales.  What is interesting about our sample is that we have variables that have larger modulation indices than the equivalent 154~MHz measurements of the \cite{Hunstead_72} variables. It would therefore be interesting to obtain higher frequency time-domain measurements of our sample, say at 400 MHz, to determine if the variability amplitudes are indeed larger at lower frequencies or if this is due to intermittency of source variability.   

The total number of variables in our sample is 15/944 or 1.6\%. These numbers are consistent with \cite{McGilchrist} who show that 1.1\% of a sample of 811 sources show fractional variability $>$15\% on time-scales of one year. Although it is difficult to compare fractional variability versus modulation index, we find no variables with a modulation index greater than 15\%. We also find a lower abundance of variability when compared with intermediate frequencies (e.g. 408~MHz) where 25\% of compact sources and 51\% of flat spectrum sources display dynamic properties (\citealt{McAdam_79}; \citealt{Fanti_81}; \citealt{spangler_1989}). All of the surveys above have flux density limits greater than 1~Jy.     

\section{Conclusions}
In this paper we report the detection of 15 candidate low-frequency variables using time-domain measurements from the MWA. The type of variability detected in this survey can be characterised as long duration ($> 5$ years) with low modulation indices ($<10\%$). In general, the variable low-frequency sky is very stable with only 1.6\% of objects displaying any variability on time-scales of years. Of our variables a large proportion (7/15) have a peaked spectrum and we conclude that low-frequency variability is more prevalent in this spectral class. If we scale up the results from this sample to our entire survey size, i.e., approximately 350,000 radio sources, we could expect an upper limit of $\sim$6000 low-frequency variables in the Southern hemisphere.  
 
We review the theory and conclude that at 154~MHz, synchrotron radiation (or intrinsic AGN variability) produces variations on time-scales of $\sim$100 years and refractive scintillation produces variations on time-scales greater than five years. The amount of variability decreases as a function of frequency, starting at 100\% at the transition frequency (typically around 5~GHz) to approximately 10\% at 150~MHz. 
Quite extreme relativistic and unrealistic boosting factors are needed to explain our results with respect to intrinsic synchrotron variations. However, we can not completely rule this out as having an impact on the variations observed, nor can we totally rule out contributions from other models such as free-free absorption.  

Based on our results, the long time-scale time-variable sky will therefore be dominated by extragalactic sources that have angular diameters approximately less than 50~mas at 154~MHz, which are fairly rare. The majority of objects observed by SKA-Low ($<$200~MHz) will therefore be non-variable. The short-timescale (minutes to days) time-variable SKA-Low sky will be dominated by objects whose angular size permits diffractive scintillation, i.e., pulsars. 

Coherent emission mechanisms and bright nearby synchrotron explosions will start to be detectable once the detection threshold drops below 1~mJy. It remains the subject of future work to thoroughly probe the short-timescale parameter space with the data collected in this survey. For example, we are still refining the model of the primary beam, which is one of the largest uncertainties in our variability analysis.    
 
The low-frequency radio sky is relatively inactive and offers a unique space to search for the counterparts to gravitational wave events. The error box currently produced by the Laser Interferometric Gravitational Wave Observatory (LIGO; \citealt{LIGO}) can be up to a few thousand square degrees on the sky. This can efficiently be covered with a few MWA pointings \citep{Kaplan_2016}. The MWA is therefore an effective instrument to search for electromagnetic signatures of merger events, not only because the field of view is so large, but also because the background rate of variable sources is low. 

At GHz frequencies the background abundance of highly variable radio sources is much higher than at low frequencies, i.e., there are approximately 3 deg$^{-2}$ (or 3000 per 1000 deg$^{2}$) at 5~GHz that show modulation indices over 50\% on month time scales \citep{Bell_2015}, which is the possible rise time for a merger event afterglow. Currently in this survey we detect no variables with such a high modulation index, therefore on a time-scale of months the background low-frequency sky will be static. This will be advantageous for identifying gravitational wave counterparts.    

\section{Acknowledgements}
TM acknowledges the support of the Australian Research Council through grant FT150100099. DLK acknowledges support by NSF grant AST-1412421. This scientific work makes use of the Murchison Radio-astronomy Observatory, operated by CSIRO. We acknowledge the Wajarri Yamatji people as the traditional owners of the Observatory site. Support for the operation of the MWA is provided by the Australian Government (NCRIS), under a contract to Curtin University administered by Astronomy Australia Limited. We acknowledge the Pawsey Supercomputing Centre which is supported by the Western Australian and Australian Governments. The NASA/IPAC Extragalactic Database (NED) is operated by the Jet Propulsion Laboratory, California Institute of Technology, under contract with the National Aeronautics and Space Administration. This research has made use of material from the Bordeaux VLBI Image Database (BVID). This work was supported by the Centre for All-sky Astrophysics (CAASTRO), an Australian Research Council Centre of Excellence (grant CE110001020).

\appendix

\label{lastpage}

\end{document}